\documentclass[12pt]{article}
\usepackage{amsmath}
\usepackage{txfonts,bm,pifont}
\usepackage{cite}
\usepackage{graphicx,xcolor}
\usepackage[dvipdfm,bookmarks=true,bookmarksnumbered=true,colorlinks=true]{hyperref}
\newtheorem{theorem}{Theorem}[section]
\newtheorem{proposition}[theorem]{Proposition}

\newtheorem{lemma}[theorem]{Lemma}
\newtheorem{remark}[theorem]{Remark}

\newcommand{\qed}{\qquad$\square$}

\numberwithin{equation}{section}
\newcommand{\utilde}[1]{\vrule depth 0pt width 0pt%
{\smash{{\mathop{#1}\limits_{\displaystyle\tilde{}}}}}}
\newcommand{\wutilde}[1]{\vrule depth 0pt width 0pt%
{\raise0.8pt\hbox{$\smash{{\mathop{#1} \limits_{\displaystyle\widetilde{}}}}$}}}
\makeatletter
\long\def\@makecaption#1#2{
 \vskip 10pt
 \setbox\@tempboxa\hbox{#1. #2}
 \ifdim \wd\@tempboxa >\hsize #1. #2\par \else \hbox
to\hsize{\hfil\box\@tempboxa\hfil}
 \fi}
\makeatother
\begin{document}
\begin{center}
\renewcommand{\baselinestretch}{1.3}\selectfont
\begin{large}
\textbf{Hypergeometric solutions to the symmetric $q$-Painlev\'e equations}
\end{large}\\[4mm]
\renewcommand{\baselinestretch}{1}\selectfont
\textrm{\large Kenji Kajiwara$^1$ and Nobutaka Nakazono$^2$}\\[2mm]
$^1$: Institute of Mathematics for Industry, Kyushu University, \\
744 Motooka, Fukuoka 819-8581, Japan.\\
kaji@imi.kyushu-u.ac.jp
\\
$^2$: School of Mathematics and Statistics, The University of Sydney, New South Wales 2006, Australia.\\
nakazono@maths.usyd.edu.au\\[2mm]
\today
\end{center}
\begin{abstract}
We consider the symmetric $q$-Painlev\'e equations derived from the
birational representation of affine Weyl groups by applying the projective
reduction and construct the hypergeometric solutions.
Moreover, we discuss continuous limits of the symmetric $q$-Painlev\'e
equations to Painlev\'e equations together with their hypergeometric
solutions.
\end{abstract}

\section{Introduction}
The discrete Painlev\'e equations, together with the Painlev\'e equations,
are now widely recognized as one of the most important families of the integrable systems
(see, for example, \cite{GR:review2004}).
Originally,  
the discrete Painlev\'e equations were discovered in the form of single second-order difference
equations and identified as the discrete analogues of the Painlev\'e
equations\cite{Brezin-Kazakov,Douglas-Shenker,Periwal-Shevitz,Fokas-Its-Kitaev,RGH:dP}.
Then they were generalized to simultaneous
first-order equations by a singularity confinement criterion\cite{GRP:SC,RGH:dP}.  A typical example
is the following equation known as a discrete Painlev\'e II equation\cite{Periwal-Shevitz,RGH:dP}:
\begin{equation}\label{sdP2:eqn}
 x_{n+1}+x_{n-1}=\frac{(an+b)x_n+c}{1-{x_n}^2},
\end{equation}
where $x_n$ is the dependent variable, $n$ is the independent variable, and $a$, $b$, $c$
$\in\mathbb{C}$ are parameters.  We note that \eqref{sdP2:eqn} was found as the equation satisfied
by the constant terms of the orthogonal polynomials on the unit circle\cite{Periwal-Shevitz}.  By
applying the singularity confinement criterion, \eqref{sdP2:eqn} is generalized to
\begin{equation}\label{adP2:eqn1}
 x_{n+1}+x_{n-1}=\frac{(an+b)x_n+c+(-1)^nd}{1-{x_n}^2},
\end{equation}
where $d$ is a parameter, with its integrability preserved. Introducing the 
dependent variables $X_n$ and $Y_n$ by 
\begin{equation}\label{dP2:specialization}
 X_n=x_{2n},\quad
 Y_n=x_{2n-1},
\end{equation}
then \eqref{adP2:eqn1} can be rewritten as
\begin{equation}\label{adP2:eqn}
 Y_{n+1}+Y_{n}=\frac{(2an+b)X_n+c+d}{1-{X_n}^2},\quad 
 X_{n+1}+X_{n}=\frac{(a(2n+1)+b)Y_{n+1}+c-d}{1-{Y_{n+1}}^2}.
\end{equation}
Equation \eqref{adP2:eqn} is known as a discrete Painlev\'e III equation since it admits a
continuous limit to the Painlev\'e III equation\cite{GNPRS:dP3,Ohta:RIMS_dP,ROSG:dP2_qP3}.
Conversely, \eqref{sdP2:eqn} can be recovered from \eqref{adP2:eqn} by putting $d=0$ and
\eqref{dP2:specialization}.  This procedure is referred to as ``{\it symmetrization}'' of
\eqref{adP2:eqn}, which comes from the terminology of the Quispel--Roberts--Thompson (QRT)
mapping\cite{QRT1,QRT2}. After this terminology, \eqref{adP2:eqn} is sometimes called the ``{\it
asymmetric}'' discrete Painlev\'e II equation, and \eqref{sdP2:eqn} is called the ``{\it
symmetric}'' discrete Painlev\'e III equation\cite{KTGR:asymmetric}.

It appears as though the symmetrization is a simple specialization on the level of the equation, but
the following problems were known: (i) According to Sakai's theory\cite{Sakai:Painleve}, 
the Painlev\'e and discrete Painlev\'e equations 
are classified by the underlying space of initial
conditions. Moreover, the discrete Painlev\'e equations arise as the birational mappings
corresponding to the translations of the affine Weyl groups associated with the space of initial
conditions. The asymmetric discrete Painlev\'e equations are characterized in this manner, however,
it was not known how to characterize the symmetric equations as the action of affine Weyl
groups. 
(ii) The Painlev\'e  and discrete Painlev\'e equations admit the particular solutions
expressible in terms of the hypergometric type functions (hypergeometric solutions) when some of the
parameters take special values (see, for example, \cite{KMNOY:hyper1,KMNOY:hyper2} and references
therein). However, the hypergeometric solutions to the symmetric discrete Painlev\'e equation cannot
be obtained by the na\"ive specialization of those to the corresponding asymmetric equation. For
example, \eqref{sdP2:eqn} has the hypergeometric solution expressible in terms of the parabolic
cylinder function (Weber function) \cite{K:dP2,KOSGR:dP2}.  On the other hand, \eqref{adP2:eqn}
admits the hypergeometric solution in terms of the confluent hypergeometric
function\cite{KNT:projective,Masuda:p5,Okamoto:p5}.  The crucial point is that although the former
function is expressed as a specialization of the latter, this specialization is not consistent with
the symmetrization.

In \cite{KNT:projective}, the mechanism of the symmetrization was investigated in detail by taking
an example of $q$-Painlev\'e equation with the affine Weyl group symmetry of type $(A_2+A_1)^{(1)}$.
Then it was shown that in general, various discrete dynamical systems of Painlev\'e type can be
obtained from elements of infinite order that are not necessarily translations in the affine Weyl
group by taking the projection on appropriate subspaces of the parameter spaces.  Such a procedure
is called a {\it projective reduction}, and the symmetrization can be understood as a kind of
projective reduction. Moreover, the above nontrivial inconsistency among the hypergeometric
solutions are explained by the factorization of the linear difference operators associated with the
three-term relation of the hypergeometric functions.

In spite of understanding the mechanism of the symmetrization or the projective reduction, 
it is still nontrivial what function will appear in the hypergeometric solutions to the symmetric
discrete Painlev\'e equations even if the hypergeometric solutions to the corresponding
asymmetric equations are known.
Since the continuous limit of the symmetric discrete Painlev\'e equation 
is different from the corresponding asymmetric one,
it is also important to consider the continuous limit of the symmetric discrete Painlev\'e equations
together with their hypergeometric solutions.

The purpose of this paper is to construct the simplest hypergeometric solutions 
to each of the symmetric $q$-Painlev\'e equations.  
In \cite{KMNOY:hyper1,KMNOY:hyper2}  the simplest hypergeometric solutions to all possible $q$-Painlev\'e equations in Sakai's list have been constructed.
Similarly, in this paper, we present the list of the simplest hypergeometric solutions to the symmetric
$q$-Painlev\'e equations reduced from the asymmetric ones with the affine Weyl group symmetry of
type $E_8^{(1)}$, $E_7^{(1)}$, $E_6^{(1)}$, $D_5^{(1)}$, $A_4^{(1)}$ and $(A_2+A_1)^{(1)}$.  We also
aim to consider the continuous limits of those symmetric $q$-Painlev\'e equations to the Painlev\'e
equations, together with their hypergeometric solutions.

This paper is organized as follows: 
in Section 2, we derive the symmetric $q$-Painlev\'e equations by applying the projective reduction to
$q$-P$((A_2+A_1)^{(1)})$, $q$-P$(A_4^{(1)})$, $q$-P$(D_5^{(1)})$, $q$-P$(E_6^{(1)})$, 
$q$-P$(E_7^{(1)})$ and $q$-P$(E_8^{(1)})$, where the $q$-Painlev\'e equation 
with the affine Weyl group of type $X$ is denoted by $q$-P($X$).
In Section 3, we construct the hypergeometric solutions to
the symmetric $q$-Painlev\'e equations derived in Section 2.
In Section 4, we discuss the continuous limits of the symmetric $q$-Painlev\'e equations
and their hypergeometric solutions.
In Section 5, we prove that hypergeometric function appearing in the hypergeometric solution
to the symmetric $q$-P$(A_4^{(1)})$ actually reduces to the Weber function by applying
the saddle point method to its integral representation.
Some concluding remarks are given in Section 6.
\section{Symmetric $q$-Painlev\'e equations}
In this section, we apply the projective reduction to the $q$-Painlev\'e
equations $q$-P$((A_2+A_1)^{(1)})$, $q$-P$(A_4^{(1)})$,
$q$-P$(D_5^{(1)})$, $q$-P$(E_6^{(1)})$, $q$-P$(E_7^{(1)})$ and
$q$-P$(E_8^{(1)})$, respectively, to obtain their symmetric forms. In
the following, we use the notations
\begin{equation}\label{eqn:basic_symbol}
\begin{split}
 \overline{t}=qt,\quad 
 \underline{t}=q^{-1}t,\quad 
 f=f(t),\quad 
 \overline{f}=f(\overline{t}),\quad 
 \underline{f}=f(\underline{t}),\\
 p^2=q,\quad  \tilde{t}=pt,\quad 
 \utilde{t}=p^{-1}t,\quad 
 \widetilde{f}=f(\tilde{t}),\quad 
 \wutilde{f}=f(\utilde{t}).
\end{split}
\end{equation}
\subsection*{Type $(A_2+A_1)^{(1)}$}
$q$-P$((A_2+A_1)^{(1)})$ is given by
\cite{KMNOY:hyper1,KMNOY:hyper2,KTGR:asymmetric,RGTT:special}
\begin{equation}\label{eqn:a2a1}
 \overline{g}g=qc^2\cfrac{1+tf}{(t+f)f}\ ,\quad
 \overline{f}f=qc^2\cfrac{1+a_2t\overline{g}}{(a_2t+\overline{g})\overline{g}}\ .
\end{equation}
The continuous limit yields the Painlev\'e III equation. 
To apply the projective reduction,
we put
\begin{equation}\label{eqn:basic_projective}
f(t)=X(t),\quad g(t)=X(p^{-1}t),\quad  a_2=p,
\end{equation} 
then \eqref{eqn:a2a1} is reduced to the symmetric $q$-P$((A_2+A_1)^{(1)})$ \cite{KNT:projective,Nakazono:A2+A1,NKT:dPII,RG:coales}
\begin{align}\label{eqn:sym_a2a1}
 \widetilde{X}\wutilde{X}=p^2c^2\cfrac{1+tX}{(t+X)X}\ .
\end{align}
The continuous limit yields the Painlev\'e II equation.

The system described in \eqref{eqn:a2a1} is a discrete dynamical system 
arising from the birational action of an element $T_1$ of the affine Weyl group of type $(A_2 + A_1)^{(1)}$, 
which is a translation on the corresponding root lattice and parameter space. 
We introduce another element of the affine Weyl group $R_1$ satisfying ${R_1}^2=T_1$. 
We note that $R_1$ is not a translation on the root lattice or on the full parameter space.
However, by taking the projection on an appropriate subspace of the parameter space, 
it becomes a translation on the subspace, which gives \eqref{eqn:sym_a2a1}.
The specialization from \eqref{eqn:a2a1} to \eqref{eqn:sym_a2a1} corresponds to this projection, 
which is an example of a projective reduction. 
We refer to \cite{KNT:projective} for details.
\subsection*{Type $A_4^{(1)}$}
$q$-P($A_4^{(1)}$)\cite{TGCR:qP4}:
\begin{equation}\label{eqn:a4}
 (\overline{g}f-1)(gf-1)=t^2\cfrac{(f+a_1)(f+{a_1}^{-1})}{f+a_2t}\ ,\quad
 (gf-1)(g\underline{f}-1)=q^{-1}t^2\cfrac{(g+a_1)(g+{a_1}^{-1})}{g+a_3t}\ .
\end{equation}
Symmetric $q$-P($A_4^{(1)}$)\cite{TGCR:qP4,RG:coales}:
\begin{equation}\label{eqn:sym_a4}
 (\widetilde{X}X-1)(X\wutilde{X}-1)=t^2\cfrac{(X+a_1)(X+{a_1}^{-1})}{X+a_2t}\ ,
\end{equation}
where
\begin{equation}
f(t)=X(t),\quad g(t)=X(p^{-1}t),\quad a_3=p^{-1}a_2.
\end{equation}
The continuous limits of \eqref{eqn:a4} and \eqref{eqn:sym_a4} yield the Painlev\'e V
and IV equations, respectively.
\subsection*{Type $D_5^{(1)}$}
$q$-P($D_5^{(1)}$)\cite{KMNOY:hyper1,RGTT:special,JS:qP6,Sakai:qp6_sol}:
\begin{equation}\label{eqn:d5}
 \overline{g}g=\cfrac{(f-q^{1/2}a_1t)(f-q^{1/2}{a_1}^{-1}t)}{(f-a_2)(f-{a_2}^{-1})},\quad
 f\underline{f}=\cfrac{(g-a_3t)(g-{a_3}^{-1}t)}{(g-a_4)(g-{a_4}^{-1})}.
\end{equation}
Symmetric $q$-P($D_5^{(1)}$)\cite{RGH:dP}:
\begin{equation}\label{eqn:sym_d5}
 \widetilde{X}\wutilde{X}=\cfrac{(X-a_1\tilde{t})(X-{a_1}^{-1}\tilde{t})}{(X-a_2)(X-{a_2}^{-1})},
\end{equation}
where
\begin{equation}
f(t)=X(t),\quad g(t)=X(p^{-1}t),\quad a_3=a_1,\quad a_4=a_2.
\end{equation}
The continuous limits of \eqref{eqn:d5} and \eqref{eqn:sym_d5} yield the
Painlev\'e VI and III equations, respectively.
\subsection*{Type $E_6^{(1)}$}
$q$-P($E_6^{(1)}$)\cite{KMNOY:hyper1,KMNOY:hyper2,RGTT:special,MSY:qPe8}:
\begin{equation}\label{eqn:e6}
\begin{cases}
 (\overline{g}f-1)(gf-1)=t^2\cfrac{(f-b_1)(f-b_2)(f-b_3)(f-b_4)}{(f-b_5t)(f-{b_5}^{-1}t)},\\[4mm]
 (gf-1)(g\underline{f}-1)=q^{-1}t^2\cfrac{(g-{b_1}^{-1})(g-{b_2}^{-1})(g-{b_3}^{-1})(g-{b_4}^{-1})}{(g-b_6u)(g-{b_6}^{-1}u)},
\end{cases}
\end{equation}
where 
\begin{equation}
 b_1b_2b_3b_4=1,\quad u=p^{-1}t.
\end{equation}
Symmetric $q$-P($E_6^{(1)}$)\cite{RGH:dP}:
\begin{equation}\label{eqn:sym_e6}
 (\widetilde{X}X-1)(X\wutilde{X}-1)=t^2\cfrac{(X-b_1)(X-{b_1}^{-1})(X-b_3)(X-{b_3}^{-1})}{(X-b_5t)(X-{b_5}^{-1}t)}.
\end{equation}
where
\begin{equation}
f(t)=X(t),\quad g(t)=X(p^{-1}t),\quad b_1b_2=1,\quad b_3b_4=1,\quad b_5b_6=1.
\end{equation}
Direct continuous limit of \eqref{eqn:e6} to the Painlev\'e equations does not exist, since it has more parameters than the Painlev\'e VI equation. 
On the other hand, \eqref{eqn:sym_e6} admits a continuous limit to the Painlev\'e V equation.
\subsection*{Type $E_7^{(1)}$}
$q$-P($E_7^{(1)}$)\cite{KMNOY:hyper1,KMNOY:hyper2,RGTT:special}:
\begin{equation}\label{eqn:e7}
  \begin{cases}
    \cfrac{(\overline{g}f-\overline{t}t)(gf-t^2)}{(\overline{g}f-1)(gf-1)}
    =\cfrac{(f-b_1t)(f-b_2t)(f-b_3t)(f-b_4t)}{(f-b_5)(f-b_6)(f-b_7)(f-b_8)},\\[4mm]
    \cfrac{(gf-t^2)(g\underline{f}-t\underline{t})}{(gf-1)(g\underline{f}-1)}
    =\cfrac{(g-{b_1}^{-1}t)(g-{b_2}^{-1}t)(g-{b_3}^{-1}t)(g-{b_4}^{-1}t)}
    {(g-{b_5}^{-1})(g-{b_6}^{-1})(g-{b_7}^{-1})(g-{b_8}^{-1})},
  \end{cases}
\end{equation}
where
\begin{equation}
  b_1b_2b_3b_4=q,\quad b_5b_6b_7b_8=1.
\end{equation}
Symmetric $q$-P($E_7^{(1)}$)\cite{GR:qP6}:
\begin{equation}\label{eqn:sym_e7}
  \cfrac{(\widetilde{X}X-\tilde{t}^2)(X\wutilde{X}-t^2)}{(\widetilde{X}X-1)(X\wutilde{X}-1)}
  =\cfrac{(X-b_1t)(X-{b_1}^{-1}\tilde{t})(X-b_3t)(X-{b_3}^{-1}\tilde{t})}{(X-b_5)(X-{b_5}^{-1})(X-b_7)(X-{b_7}^{-1})}.
\end{equation}
where
\begin{equation}
  f(t)=X(t),\quad g(t)=X(p^{-1}t),\quad b_1b_2=p,\quad b_3b_4=p,\quad b_5b_6=1,\quad b_7b_8=1.
\end{equation}
Direct continuous limit of \eqref{eqn:e7} to the Painlev\'e equations does
not exist, while that of \eqref{eqn:sym_e7} yields the Painlev\'e VI equation.
\subsection*{Type $E_8^{(1)}$}
$q$-P($E_8^{(1)}$)\cite{KMNOY:hyper1,KMNOY:hyper2,RGTT:special,MSY:qPe8,ORG:8para_dP}:
\begin{equation}\label{eqn:e8}
\begin{cases}
 \cfrac{(\overline{g}\,\overline{u}t-f)(gut-f)-(\overline{u}^2t^2-1)(u^2t^2-1)}
  {(\overline{u}^{-1}t^{-1}\overline{g}-f)(u^{-1}t^{-1}g-f)-(1-\overline{u}^{-2}t^{-2})(1-u^{-2}t^{-2})}
 =\cfrac{P(f,t,m_1,\cdots,m_7)}{P(f,t^{-1},m_7,\cdots,m_1)},\\[4mm]
 \cfrac{(fut-g)(\underline{f}u\underline{t}-g)-(u^2t^2-1)(u^2\underline{t}^2-1)}
  {(u^{-1}t^{-1}f-g)(u^{-1}\underline{t}^{-1}\underline{f}-g)-(1-u^{-2}t^{-2})(1-u^{-2}\underline{t}^{-2})}
 =\cfrac{P(g,u,m_7,\cdots,m_1)}{P(g,u^{-1},m_1,\cdots,m_7)},
\end{cases}
\end{equation}
where 
\begin{equation}
\begin{split}
 P(f,t,m_1,\cdots,m_7)
 =&f^4-m_1tf^3+(m_2t^2-3-t^8)f^2\\
 &+(m_7t^7-m_3t^3+2m_1t)f+t^8-m_6t^6+m_4t^4-m_2t^2+1.
\end{split}
\end{equation}
Here $m_j$ $(j=1,\cdots,8)$ are the elementary symmetric functions of $j$-th degree in $b_k$ $(k=1,\cdots,8)$,
$m_8=1$, and $u=p^{-1}t$.
Symmetric $q$-P($E_8^{(1)}$):
\begin{equation}\label{eqn:sym_e8}
 \cfrac{(\tilde{t}t\widetilde{X}-X)(t\utilde{t}\wutilde{X}-X)-(\tilde{t}^2t^2-1)(t^2\utilde{t}^2-1)}
  {(\tilde{t}^{-1}t^{-1}\widetilde{X}-X)(t^{-1}\utilde{t}^{-1}\wutilde{X}-X)
  -(\tilde{t}^{-2}t^{-2}-1)(t^{-2}\utilde{t}^{-2}-1)}
 =\cfrac{P(X,t,m_1,m_2,m_3,m_4,m_3,m_2,m_1)}{P(X,t^{-1},m_1,m_2,m_3,m_4,m_3,m_2,m_1)}.
\end{equation}
where
\begin{equation}
f(t)=X(t),\quad g(t)=X(p^{-1}t),\quad m_i=m_{8-i}\quad(i=1,2,3).
\end{equation}
Direct continuous limits of \eqref{eqn:e8} and \eqref{eqn:sym_e8} to the Painlev\'e equations are
not known.
\section{Hypergeometric solutions to the symmetric $q$-Painlev\'e equations}
In this section, we construct the hypergeometric solutions to the
symmetric $q$-Painlev\'e equations.
We use the following conventions of $q$-analysis\cite{Gasper-Rahman:BHS}.
The basic hypergeometric series ${}_s\varphi_r$ is defined by
\begin{equation}
 {}_s\varphi_r\left(\begin{matrix}a_1,\cdots,a_s\\b_1,\cdots,b_r\end{matrix};q,z\right)
 =\sum_{n=0}^{\infty}\cfrac{(a_1,\cdots,a_s;q)_n}{(b_1,\cdots,b_r;q)_n(q;q)_n}
 \begin{bmatrix}(-1)^nq^{n(n-1)/2}\end{bmatrix} ^{1+r-s}z^n,
\end{equation}
where
\begin{equation}
 (a_1,\cdots,a_s;q)_n=\prod_{i=1}^s(a_i;q)_n,\quad  (a;q)_k=\prod_{i=1}^{k}(1-aq^{i-1}),
\end{equation}
are the $q$-shifted factorials. The following special case of the basic hypergeometric series is known
as the very-well-poised basic hypergeometric series
\begin{equation}
 {}_{r+1}W_r(a_1;a_4,a_5,\cdots,a_{r+1};q,z)
 ={}_{r+1}\varphi_r\left(\begin{matrix}a_1,q{a_1}^{1/2},-q{a_1}^{1/2},a_4,\cdots,a_{r+1}\\
 {a_1}^{1/2},-{a_1}^{1/2},qa_1{a_4}^{-1},\cdots,qa_1{a_{r+1}}^{-1}\end{matrix};q,z\right).
\end{equation}
We also use the Jacobi theta function
\begin{equation}
 \Theta(a;q)=(a;q)_\infty(qa^{-1};q)_\infty,
\end{equation}
which satisfies the $q$-difference equation
\begin{equation}
 \Theta(qa;q)=-a^{-1}\Theta(a;q).
\end{equation}
\subsection{Symmetric $q$-P($(A_2+A_1)^{(1)}$)}
\begin{proposition}\label{prop:hyper_sym_a2a1}
Symmetric $q$-{\rm P}$((A_2+A_1)^{(1)})$ \eqref{eqn:sym_a2a1} 
admits the hypergeometric solution
\begin{equation}\label{eqn:riccati_3term_a2a1}
 \widetilde{X}=p^{1/2}\cfrac{\widetilde{G}}{G}\ ,
\end{equation}
\begin{equation}\label{eqn:sol_a2a1}
 G=Ae^{(\pi i/2)(\log t/\log p)}
 {}_1\varphi_1 \left(\begin{matrix}0\\-p\end{matrix};p,ip^{3/2}t\right)
 +Be^{(-\pi i/2)(\log t/\log p)}
 {}_1\varphi_1 \left(\begin{matrix}0\\-p\end{matrix};p,-ip^{3/2}t\right),
\end{equation}
with
\begin{equation}\label{eqn:condition_a2a1}
 c=1.
\end{equation}
Here, $A$ and $B$ are quasi-constants satisfying $ A(t)=A(pt)$ and $B(t)=B(pt)$, respectively.
\end{proposition}
\noindent{\bf Proof.} 
Substituting 
\begin{equation}
 \widetilde{X}=\cfrac{P(t)X+Q(t)}{R(t)X+S(t)},\quad
 \wutilde{X}=\cfrac{-S(p^{-1}t)X+Q(p^{-1}t)}{R(p^{-1}t)X-P(p^{-1}t)},
\end{equation}
in \eqref{eqn:sym_a2a1},
we find that \eqref{eqn:sym_a2a1} admits a specialization to the discrete Riccati equation 
\begin{equation}\label{eqn:riccati_a2a1}
 \widetilde{X}=-p\cfrac{1+tX}{X},
\end{equation}
when $c=1$. Then putting as \eqref{eqn:riccati_3term_a2a1}),
\eqref{eqn:riccati_a2a1} is linearized to the three-term relation for
$G$
\begin{equation}\label{eqn:3term_a2a1}
 \widetilde{\widetilde{G}}+p^{3/2}t~\widetilde{G}+G=0.
\end{equation}
Moreover, substituting the power series expression
\begin{equation}\label{eq:sya2a1_sen}
 G=\displaystyle\sum^{\infty}_{n=0}C_nt^{n+\rho},\quad (\rho,~C_n\in\mathbb{C}),
\end{equation}
into \eqref{eqn:3term_a2a1}), we obtain \eqref{eqn:sol_a2a1}.\hfil\qed

In the following, we present the hypergeometric solutions to other
cases. Since the results are verified by direct calculations, we omit
the proof showing only the discrete Riccati equations and the linearized
three-term relations. We also assume that $A$ and $B$ are
quasi-constants.
\subsection{Symmetric $q$-P($A_4^{(1)}$)}
\begin{proposition}\label{prop:hyper_sym_a4}
Symmetric $q$-{\rm P}$(A_4^{(1)})$ \eqref{eqn:sym_a4} admits the hypergeometric solution
\begin{equation}\label{eqn:riccati_3term_a4}
 \widetilde{X}=\cfrac{F}{\widetilde{F}}\ ,
\end{equation}
\begin{equation}\label{eqn:sol_a4}
 F=(p{a_2}^{-1}t;p)_\infty\left(
 A~ {}_2\varphi_1\left(\begin{matrix}{a_2}^2,0\\-p\end{matrix};p,p{a_2}^{-1}t\right)
 +B~\ (-1)^{\log t/\log p}
 {}_2\varphi_1\left(\begin{matrix}-{a_2}^2,0\\-p\end{matrix};p,p{a_2}^{-1}t\right)
 \right),
\end{equation}
with
\begin{equation}\label{eqn:condition_a4}
 a_1=p^{-1}{a_2}^2.
\end{equation}
\end{proposition}
The discrete Riccati equation and the linearized three-term relation are given by
\begin{equation}\label{eqn:riccati_a4}
 \widetilde{X}=\cfrac{1-p{a_2}^{-1}t}{X+a_2t},
\end{equation}
and 
\begin{equation}\label{eqn:3term_a4}
 (p{a_2}^{-1}t-1)\widetilde{F}+a_2tF+\wutilde{F}=0,
\end{equation}
respectively.

\subsection{Symmetric $q$-P($D_5^{(1)}$)}
For convenience, we set
\begin{equation}
 X=-iW,\quad a_{1}=ip^{\nu_{1}/2},\quad a_{2}=ip^{\nu_{2}/2}.
\end{equation}
Symmetric $q$-P$(D_{5}^{(1)})$  \eqref{eqn:sym_d5} can be rewritten as
\begin{equation}\label{eqn:sym_d5_2}
 \widetilde{W}\wutilde{W}=-\cfrac{(W+p^{\nu_{1/2}}\tilde{t}\,)(W-p^{-\nu_{1}/2}\tilde{t}\,)}{(W+p^{\nu_{2}/2})(W-p^{-\nu_{2}/2})}~.
\end{equation}
\begin{proposition}[\cite{KOS:dp3,KS:q2DTL}]\label{prop:hyper_sym_d5}
Symmetric $q$-{\rm P}$(D_5^{(1)})$ \eqref{eqn:sym_d5_2} admits the hypergeometric solution
\begin{equation}
 W=\cfrac{\widetilde{G}}{G}-p^{\nu_{2}/2}\ ,
\end{equation}
\begin{equation}\label{eqn:sol_d5}
 G=A~J_{\nu_{2}}^{(1)}(2ip^{3/4}t^{1/2};p)+B~J_{-\nu_{2}}^{(1)}(2ip^{3/4}t^{1/2};p),
\end{equation}
with
\begin{equation}\label{eqn:condition_d5}
 \nu_1=\nu_{2}+1.
\end{equation}
Here,  $J_{\nu}^{(1)}(x;q)$ is Jackson's $q$-Bessel fuction\cite{Gasper-Rahman:BHS}
\begin{equation}
  J_{\nu}^{(1)}(x;q)
  =\frac{(q^{\nu+1};q)_{\infty}}{(q;q)_{\infty}}~\left(\frac{x}{2}\right)^{\nu}~{}_2\varphi_1\left(\begin{matrix}0,0\\q^{\nu+1}\end{matrix};q,-\frac{x^{2}}{4}\right).
\end{equation}
\end{proposition}
The discrete Riccati equation and the linearized three-term relation are given by
\begin{equation}\label{eqn:riccati_d5}
 \widetilde{W}=p^{-\nu_{2}/2}\cfrac{W+p^{(\nu_{2}+1)/2}\tilde{t}}{W+p^{\nu_{2}/2}},
\end{equation}
and
\begin{equation}\label{eqn:3term_d5}
 \widetilde{\widetilde{G}}-(p^{\nu_{2}/2}+p^{-\nu_{2}/2})\widetilde{G} +(1-p^{1/2}\tilde{t})G=0,
\end{equation}
respectively.

\subsection{Symmetric $q$-P($E_6^{(1)}$)}
\begin{proposition}\label{prop:hyper_sym_e6}
Symmetric $q$-{\rm P}$(E_6^{(1)})$ \eqref{eqn:sym_e6} admits the hypergeometric solution
\begin{equation}
 X=({b_1}^{-1}b_5t-1)\cfrac{\widetilde{G}}{G}+b_5 t,
\end{equation}
\begin{equation}\label{eqn:sol_e6}
 G=A~{}_2\varphi_1\left(\begin{matrix}{b_1}^{-1}{b_5}^2,-{b_1}^{-1}\\-p\end{matrix}
  ;p,b_1{b_5}^{-1}\tilde{t}\right)
 +B\cfrac{\Theta(t;p)}{\Theta(-t;p)}~
 {}_2\varphi_1\left(\begin{matrix} -{b_1}^{-1}{b_5}^2,{b_1}^{-1}\\ -p\end{matrix}
  ;p,b_1{b_5}^{-1}\tilde{t}\right),
\end{equation}
with
\begin{equation}\label{eqn:condition_e6}
 {b_5}^2=pb_1b_3.
\end{equation}
\end{proposition}
The discrete Riccati equation and the linearized three-term relation are given by
\begin{equation}\label{eqn:riccati_3term_e6}
 \widetilde{X}=\cfrac{b_1\tilde{t}X+b_1b_5-(p{b_1}^2+{b_5}^2)t}{b_1b_5(X-b_5t)},
\end{equation}
\begin{equation}\label{eqn:3term_e6}
 (b_5t-b_1)b_5\widetilde{G}+b_1({b_5}^2-1)tG-b_1(b_1t-b_5)\wutilde{G}=0,
\end{equation}
respectively.

\subsection{Symmetric $q$-P($E_7^{(1)}$)}
\begin{proposition}\label{prop:hyper_sym_e7}
Symmetric $q$-{\rm P}$(E_7^{(1)})$ \eqref{eqn:sym_e7} admits the hypergeometric solution
\begin{equation}\label{eqn:riccati_3term_e7}
 \cfrac{X-{b_1}^{-1}\tilde{t}}{X-b_5}
 =\frac{(b_3b_5\tilde{t}-1)}{(b_5^2-1)(\tilde{t}-b_3b_5)}
 \left((b_5\tilde{t}-{b_1}^{-1})\cfrac{\widetilde{G}}{G}+\tilde{t}-{b_1}^{-1}b_5\right),
\end{equation}
\begin{equation}\label{eqn:sol_e7_0}
\begin{split}
 &G=AG_1+BG_2,\\
 &G_1=\cfrac{(b_1 b_5 \tilde{t},-b_3 \tilde{\tilde{t}};p)_{\infty }(b_1 b_5 b_3)^{\log t/\log p}}
  {({b_1}^{-1} {b_5}^{-1}\tilde{\tilde{t}},-{b_3}^{-1}\tilde{t};p)_{\infty }}\\
 &\qquad\times{}_8W_7\left(-{b_3}^2 b_5;p^{1/2}b_3,-p^{1/2}b_3,-b_1 b_3 b_5,b_3
  b_5\tilde{t},b_3 b_5\tilde{t}^{-1};p,p{b_1}^{-1}{b_3}^{-1}{b_5}^{-1}\right),\\
 &G_2=\cfrac{(\tilde{\tilde{t}}\tilde{t},-p^{1/2}{b_1}^{-1}{b_3}^{-1}{b_5}^{-1}\tilde{\tilde{t}},
   p^{1/2}{b_1}^{-1}{b_3}^{-1}{b_5}^{-1}\tilde{\tilde{t}},{b_3}^{-1}\tilde{\tilde{t}},
   {b_1}^{-1}\tilde{\tilde{t}};p)_{\infty }(-b_5)^{\log t/\log p}}
  {({b_1}^{-1}{b_5}^{-1}\tilde{\tilde{t}},p^{1/2}\tilde{t},-p^{1/2}\tilde{t},{b_3}^{-1}{b_5}^{-1}
   \tilde{\tilde{t}},{b_1}^{-1}{b_3}^{-1}{b_5}^{-1}\tilde{\tilde{t}}^2;p)_{\infty }}\\
 &\qquad\times
 {}_8W_7\left({b_1}^{-1}{b_3}^{-1}{b_5}^{-1}\tilde{\tilde{t}}\tilde{t};{b_1}^{-1}{b_5}^{-1}\tilde{\tilde{t}},
  p^{1/2}\tilde{t},-p^{1/2}\tilde{t},{b_3}^{-1}{b_5}^{-1}\tilde{\tilde{t}},
  p{b_1}^{-1}{b_3}^{-1}{b_5}^{-1};p,-b_5\right),
\end{split}
\end{equation}
with
\begin{equation}\label{eqn:condition_e7}
 b_1b_3b_5b_7=1.
\end{equation}
\end{proposition}
The discrete Riccati equation and the linearized three-term relation are given by
\begin{equation}\label{eqn:riccati_e7}
 \cfrac{\widetilde{X}X-\tilde{t}^2}{\widetilde{X}X-1}
 =\cfrac{(X-{b_1}^{-1}\tilde{t})(X-{b_3}^{-1}\tilde{t})}{(X-b_5)(X-{b_1}^{-1}{b_3}^{-1}{b_5}^{-1})},
\end{equation}
\begin{equation}\label{eqn:3term_1_e7}
 \cfrac{(b_1b_5\tilde{t}-1)(b_3b_5\tilde{t}-1)}{(b_1b_3b_5-1)(1+b_5)}(\widetilde{G}-G)
 +(\tilde{t}^2-1)G+\cfrac{(\tilde{t}-b_1b_5)(\tilde{t}-b_3b_5)}{(b_1b_3b_5-1)(1+b_5)}(G-\wutilde{G})=0,
\end{equation}
respectively.  To obtain the solution of \eqref{eqn:3term_1_e7}, the following proposition is useful:
\begin{proposition}[\cite{IR:askey}]\label{prop:3term_askey}
The three-term relation for $G=AG_1+BG_2$, where
\begin{align}
 G_1=&\cfrac{(az^{-1})^{\log u/\log p}(abu,acu,adu,bcdz^{-1}u)_{\infty }}
  {(bcu,bdu,cdu,azu)_{\infty }}\notag\\
 &\times{}_8W_7(p^{-1}bcdz^{-1};bz^{-1},cz^{-1},dz^{-1},p^{-1}abcdu,u^{-1};p,pa^{-1}z),\\
 G_2=&\cfrac{(az)^{\log u/\log p}(abcdu^2,pbzu,pczu,pdzu,bcdzu)_{\infty }}
  {(bcu,bdu,cdu,pu,pbcdzu^2)_{\infty }}\notag\\
 &\times{}_8W_7(bcdzu^2;bcu,bdu,cdu,pu,pa^{-1}z;p,az),
\end{align}
is given by
\begin{equation}
\begin{split}
 &\cfrac{(1-abu)(1-acu)(1-adu)(1-p^{-1}abcdu)}{a(1-p^{-1}abcdu^2)(1-abcdu^2)}~(G(pu)-G(u))
 +(a+a^{-1}-z-z^{-1})G(u)\\
 &-\cfrac{a(1-p^{-1}bcu)(1-p^{-1}bdu)(1-p^{-1}cdu)(1-u)}{(1-p^{-2}abcdu^2)(1-p^{-1}abcdu^2)}
 ~(G(u)-G(p^{-1}u))=0.
\end{split}
\end{equation}
Here $A$ and $B$ are quasi-constants with respect to $u$.
\end{proposition}
Substituting
\begin{equation}
\begin{split}
 &a=i{b_1}^{1/2}{b_3}^{1/2}b_5,\quad
 b=ip^{1/2}{b_1}^{-1/2}{b_3}^{1/2},\quad
 c=-ip^{1/2}{b_1}^{-1/2}{b_3}^{1/2},\\
 &d=-i{b_1}^{1/2}{b_3}^{1/2}b_5,\quad
 z=i{b_1}^{-1/2}{b_3}^{-1/2},\quad
 u={b_3}^{-1}{b_5}^{-1}t, 
 \end{split}
\end{equation}
in Proposition \ref{prop:3term_askey}, we obtain the solution of \eqref{eqn:3term_1_e7}.

For later convenience, we rewrite the solution in the following manner:
\begin{equation}
 F=
 \cfrac{({b_1}^{-1} {b_5}^{-1}\tilde{\tilde{t}},-{b_3}^{-1}\tilde{t};p)_{\infty }}
 {(b_1 b_5 \tilde{t},-b_3 \tilde{\tilde{t}};p)_{\infty }(b_1 b_5 b_3)^{\log t/\log p}}~
 G.
\end{equation}
Then Proposition \ref{prop:hyper_sym_e7} is rephrased as follows:
\begin{proposition}
Symmetric $q$-{\rm P}$(E_7^{(1)})$ \eqref{eqn:sym_e7} admits the hypergeometric solution
\begin{equation}\label{eqn:riccati_3term_2_e7}
 \frac{X-{b_1}^{-1}\tilde{t}}{X-b_5}
 =\frac{b_3b_5\tilde{t}-1}{b_1({b_5}^2-1)(\tilde{t}-b_3b_5)}
 \left[\frac{(\tilde{\tilde{t}}-b_1b_5)(\tilde{t}+b_3)}
   {b_3\tilde{\tilde{t}}+1}\cfrac{\widetilde{F}}{F}
  +b_1\tilde{t}-b_5\right],
\end{equation}
\begin{equation}\label{eqn:sol_e7}
\begin{split}
 &F=AF_1+BF_2,\\
 &F_1=
 {}_8W_7\left(-{b_3}^2 b_5;p^{1/2}b_3,-p^{1/2}b_3,-b_1 b_3 b_5,
 b_3b_5\tilde{t},b_3 b_5\tilde{t}^{-1};p,p{b_1}^{-1}{b_3}^{-1}{b_5}^{-1}\right),\\
 &F_2=\cfrac{(-{b_3}^{-1}\tilde{t},\tilde{\tilde{t}}\tilde{t},
 -p^{1/2}{b_1}^{-1}{b_3}^{-1}{b_5}^{-1}\tilde{\tilde{t}},
 p^{1/2}{b_1}^{-1}{b_3}^{-1}{b_5}^{-1}\tilde{\tilde{t}},
 {b_3}^{-1}\tilde{\tilde{t}},{b_1}^{-1}\tilde{\tilde{t}};p)_{\infty }}
 {(b_1 b_5 \tilde{t},-b_3 \tilde{\tilde{t}},p^{1/2}\tilde{t},
 -p^{1/2}\tilde{t},{b_3}^{-1}{b_5}^{-1}
 \tilde{\tilde{t}},{b_1}^{-1}{b_3}^{-1}{b_5}^{-1}\tilde{\tilde{t}}^2;p)_{\infty }
 (-b_1b_3)^{\log t/\log p}}\\
 &\hspace{2em}\times
 {}_8W_7\left({b_1}^{-1}{b_3}^{-1}{b_5}^{-1}\tilde{\tilde{t}}\tilde{t};
 {b_1}^{-1}{b_5}^{-1}\tilde{\tilde{t}},
 p^{1/2}\tilde{t},-p^{1/2}\tilde{t},
 {b_3}^{-1}{b_5}^{-1}\tilde{\tilde{t}},p{b_1}^{-1}{b_3}^{-1}{b_5}^{-1};p,-b_5\right).
\end{split}
\end{equation}
when
\begin{equation}\label{eqn:condition_e7_2}
 b_1b_3b_5b_7=1.
\end{equation}
\end{proposition}
We note that the linear three-term relation for $F$ is given by
\begin{align}\label{eqn:3term_2_e7}
 &\cfrac{(b_1b_5\tilde{t}-1)(b_3b_5\tilde{t}-1)}{(b_1b_3b_5-1)(1+b_5)}
  \left(\cfrac{(b_1b_5-\tilde{\tilde{t}})(b_3+\tilde{t})}
  {(1-b_1 b_5 \tilde{t})(1+b_3 \tilde{\tilde{t}})}\widetilde{F}-F\right)
 +(\tilde{t}^2-1)F\notag\\
 &\quad+\cfrac{(\tilde{t}-b_1b_5)(\tilde{t}-b_3b_5)}{(b_1b_3b_5-1)(1+b_5)}
  \left(F
  -\cfrac{(1-b_1 b_5 t)(1+b_3 \tilde{t})}{(b_1b_5-\tilde{t})(b_3+t)}\wutilde{F}\right)
 =0.
\end{align}
\subsection{Symmetric $q$-P($E_8^{(1)}$)}
\begin{proposition}\label{prop:hyper_sym_e8}
Symmetric $q$-{\rm P}$(E_8^{(1)})$ \eqref{eqn:sym_e8} admits the hypergeometric solution
\begin{equation}
 \cfrac{X-\beta_3}{X-\beta_1}
 =\frac{(\alpha_3-\alpha_5)(\beta_1-\beta_5)}{(\alpha_1-\alpha_5)(\beta_3-\beta_5)}\left[(\alpha_1-\alpha_3)
 \cfrac{\widetilde{G}}{G}+\alpha_3-\tilde{\beta}_1\right],
\end{equation}
\begin{equation}\label{eqn:sol_e8}
\begin{split}
 &G=AG_1+BG_2,\\
 &G_1=\cfrac{(b_1b_3\tilde{t}^2,b_1b_5\tilde{t}^2,-\tilde{t}^2,{b_1}^{-1}{b_3}^{-1}{b_5}^{-1}\tilde{t}^2;p^2)_{\infty }}
  {(-\tilde{t}^2,{b_1}^{-1}{b_5}^{-1}\tilde{t}^2,{b_1}^{-1}{b_3}^{-1}\tilde{t}^2,b_1b_3b_5\tilde{t}^2;p^2)_{\infty }}
 (-b_1)^{\log t/\log p}\\
 &\quad\times{}_8W_7(p^{-2}{b_1}^{-1}{b_3}^{-2}{b_5}^{-2};{b_5}^{-1},{b_3}^{-1},-{b_1}^{-1}{b_3}^{-1}{b_5}^{-1},
  {b_3}^{-1}{b_5}^{-1}t^2,{b_3}^{-1}{b_5}^{-1}\tilde{t}^{-2};p^2,-p^2{b_1}^{-1}),\\
 &G_2=\cfrac{(\tilde{t}^4,-b_3\tilde{\tilde{t}}^2,-b_5\tilde{\tilde{t}}^2,
   {b_1}^{-1}\tilde{\tilde{t}}^2,-{b_1}^{-1}{b_3}^{-1}{b_5}^{-1}\tilde{t}^2;p^2)_{\infty }}
  {(-\tilde{t}^2,{b_1}^{-1}{b_5}^{-1}\tilde{t}^2,{b_1}^{-1}{b_3}^{-1}\tilde{t}^2,b_3b_5\tilde{\tilde{t}}^2,
   -{b_1}^{-1}\tilde{\tilde{t}}^2\tilde{t}^2;p^2)_{\infty }}
  {b_1}^{\log t/\log p}\\
 &\quad\times{}_8W_7(-{b_1}^{-1}\tilde{t}^4;-\tilde{t}^2,{b_1}^{-1}{b_5}^{-1}\tilde{t}^2,
  {b_1}^{-1}{b_3}^{-1}\tilde{t}^2,b_3b_5\tilde{\tilde{t}}^2,-p^2{b_1}^{-1};p^2,b_1),
\end{split}
\end{equation}
with
\begin{equation}
 b_1b_3b_5b_7=p^{-2}.
\end{equation}
Here, $\alpha_i$ and $\beta_i$ are given by
\begin{equation}
 \alpha_i=b_i\tilde{t}+\cfrac{1}{b_i\tilde{t}}\ ,\quad 
 \beta_i=\cfrac{t}{b_i}+\cfrac{b_i}{t}.
\end{equation}
\end{proposition}
The discrete Riccati equation and the linearized three-term relation are given by
\begin{equation}\label{eqn:riccati_e8}
 \widetilde{X}
 =\cfrac{X\begin{vmatrix}1&\alpha_1&\alpha_1\beta_1\\1&\alpha_3&\alpha_3\beta_3\\1&\alpha_5&\alpha_5\beta_5\end{vmatrix}
 +\begin{vmatrix}\alpha_1&\beta_1&\alpha_1\beta_1\\\alpha_3&\beta_3&\alpha_3\beta_3\\
   \alpha_5&\beta_5&\alpha_5\beta_5\end{vmatrix}}
 {X\begin{vmatrix}1&\alpha_1&\beta_1\\1&\alpha_3&\beta_3\\1&\alpha_5&\beta_5\end{vmatrix}
 +\begin{vmatrix}1&\beta_1&\alpha_1\beta_1\\1&\beta_3&\alpha_3\beta_3\\1&\beta_5&\alpha_5\beta_5\end{vmatrix}},
\end{equation}
\begin{equation}\label{eqn:3term_e8}
\begin{split}
 &\cfrac{(b_1b_3\tilde{t}^2-1)(b_1b_5\tilde{t}^2-1)(t^2-b_3b_5)}{(\tilde{t}^2-1)(\tilde{t}^2t^2-1)}(\widetilde{G}-G)
 +b_3b_5({b_1}^2-1)G\\
 &\quad+\cfrac{(t^2-b_1b_3)(t^2-b_1b_5)(b_3b_5\tilde{t}^2-1)}{(t^2-1)(\tilde{t}^2t^2-1)}(G-\wutilde{G})=0,
\end{split} 
\end{equation}
respectively. We note that \eqref{eqn:sol_e8} is obtained by substituting
\begin{equation}
 a=-ib_1,\quad
 b=i{b_5}^{-1},\quad
 c=i{b_3}^{-1},\quad
 d=-i{b_1}^{-1}{b_3}^{-1}{b_5}^{-1},\quad
 z=i,\quad
 u=b_3b_5\tilde{t}^2,
\end{equation}
in Proposition \ref{prop:3term_askey}.
\section{Continuous limits}
In this section, we discuss the continuous limits of the symmetric
$q$-Painlev\'e equations and their hypergeometric solutions. We introduce the hypergeometric series 
${}_sF_r$ defined by
\begin{equation}
 {}_sF_r\left(\begin{matrix}a_1,\cdots,a_s\\b_1,\cdots,b_r\end{matrix};z\right)
 =\sum_{n=0}^{\infty}\cfrac{(a_1)_n\cdots(a_s)_n}{(b_1)_n\cdots(b_r)_nn!}~z^n,
\end{equation}
where $(a)_n=a(a+1)\dots(a+n-1)$.
\subsection{Symmetric $q$-P($(A_2+A_1)^{(1)}$)}
It is known that symmetric $q$-{\rm P}$((A_2+A_1)^{(1)})$ \eqref{eqn:sym_a2a1} yields
the Painlev\'e II equation
\begin{equation}\label{eqn:p2}
 z''=2z^3+sz+\alpha,
\end{equation}
where $z'={\rm d}z/{\rm d}s$, by putting
\begin{equation}\label{eqn:param_continuous_limit_a2a1}
 t=-2e^{-s\epsilon^2/4-\alpha\epsilon^3/4},\quad
 c=e^{(1+2\alpha)\epsilon^3/4},\quad
 X=e^{-s\epsilon^2/4+\alpha\epsilon^3/4}(1+\epsilon z),\quad
 p=e^{-\epsilon^3/4},
\end{equation} 
and taking the limit $\epsilon\to+0$ \cite{RG:coales}. The limit of the hypergeometric solution is
given as follows:
\begin{proposition}
Under the parametrization \eqref{eqn:param_continuous_limit_a2a1}, the hypergeometric solution in Proposition \ref{prop:hyper_sym_a2a1} 
is reduced to the following solution to \eqref{eqn:p2}:
\begin{equation}
 z=\cfrac{G'}{G}~,
\end{equation}
\begin{equation}\label{eqn:sol_p2}
 G=A~{\rm Ai}(2^{-1/3}se^{-\pi i/3})+B~{\rm Ai}(2^{-1/3}se^{\pi i/3}),
\end{equation}
with
\begin{equation}
 \alpha=-\cfrac{1}{2}.
\end{equation}
Here $A$ and $B$ are constants, and ${\rm Ai}(x)$ is the Airy function
\begin{equation}
 {\rm Ai}(x)=\cfrac{1}{2\pi i}\int_{-\infty}^\infty e^{iu^3/3+ixu}du.
\end{equation}
\end{proposition}
We note that \eqref{eqn:riccati_a2a1} and \eqref{eqn:3term_a2a1} are reduced to the following equations
\begin{equation}
 z'=-z^2-\cfrac{s}{2},
\end{equation}
\begin{equation}
 G''+\cfrac{s}{2}~G=0,
\end{equation}
respectively. We also note that \eqref{eqn:sol_p2} is obtained from
\eqref{eqn:sol_a2a1} by using the following proposition:
\begin{proposition}[\cite{HKW:A1A1}]\label{prop:continuous_airy}
With the substitutions
\begin{equation}
 q=e^{-\delta^3/2},\quad
 T=-2ie^{-(u/2)\delta^2},
\end{equation}
as $\delta\to+0$, it follows that
\begin{align}
 &{}_1\varphi_1 \left(\begin{matrix}0\\-q\end{matrix};q,qT\right)
 =2\pi^{1/2}\delta^{-1/2}e^{\log2(\pi i/\delta^3)-(\pi i/2\delta)u-\pi i/12}
 \left[{\rm Ai}(ue^{-\pi i/3})+O(\delta^2)\right],\\
 &{}_1\varphi_1 \left(\begin{matrix}0\\-q\end{matrix};q,-qT\right)
 =2\pi^{1/2}\delta^{-1/2}e^{-\log2(\pi i/\delta^3)+(\pi i/2\delta)u+\pi i/12}
 \left[{\rm Ai}(ue^{\pi i/3})+O(\delta^2)\right],
\end{align}
for $u$ in any compact domain of $\mathbb{C}$.
\end{proposition}

\subsection{Symmetric $q$-P($A_4^{(1)}$)}
Putting
\begin{equation}\label{eqn:condition_limit_a4}
 t=2ie^{is\epsilon-3\alpha\epsilon^2/2},\quad
 a_1=-e^{-2^{-1/2}i\beta^{1/2}\epsilon^2},\quad
 a_2=ie^{-\alpha\epsilon^2/2},\quad
 X=1-i\epsilon z,\quad
 p=e^{-\epsilon^2},
\end{equation}
and taking the limit $\epsilon\to 0$, symmetric $q$-{\rm P}$(A_4^{(1)})$ \eqref{eqn:sym_a4} yields the
Painlev\'e IV equation\cite{{RG:coales}}
\begin{equation}
 z''=\cfrac{(z')^2}{2z}+\cfrac{3}{2}z^3+4sz^2+2(s^2-\alpha)z+\cfrac{\beta}{z}.
\end{equation}
\begin{proposition}\label{prop:A4_sol}
Under the parametrization \eqref{eqn:condition_limit_a4}, 
the hypergeometric solution in Proposition \ref{prop:hyper_sym_a4} is reduced to
\begin{equation}
 z=\frac{F'}{F},
\end{equation}
\begin{equation}\label{eqn:sol_p4}
 F=A~e^{-s^2}D_{-\alpha}(2^{1/2}s)+B~D_{\alpha-1}(2^{1/2}is),\quad s\not\in \mathbb{R}, \sqrt{-1}\mathbb{R},
\end{equation}
with 
\begin{equation}
 \alpha=1+2^{-1/2}i\beta^{1/2}.
\end{equation}
Here $A$ and $B$ are constants, and $D_\lambda(x)$ is the Weber function
\begin{equation}
 D_\lambda(x)=\int_\gamma e^{-u^2/2-xu}u^{-\lambda-1}du,
\end{equation}
where the path of integration $($denote it by $\gamma$$)$ runs from $-\infty$ to $+\infty$
so that $u=0$ lies to the right of the path. 
\end{proposition}
\begin{remark}
The hypergeometric solution to the Painlev\'e IV equation given in Proposition \ref{prop:A4_sol} is consistent with the solution appeared in \cite{Noumi:book,Okamoto:p24}.
\end{remark}
Note that \eqref{eqn:riccati_a4} and \eqref{eqn:3term_a4} are reduced to the following equations
\begin{equation}
 z'=-z^2-2sz+2(\alpha-1),
\end{equation}
\begin{equation}
 F''+2sF'+2(1-\alpha)F=0.
\end{equation}
respectively.
We also note that \eqref{eqn:sol_p4} is obtained from \eqref{eqn:sol_a4} by using the following lemmas:
\begin{lemma}\label{lemma:continuous_weber_1}
With the substitutions \eqref{eqn:condition_limit_a4}, it follows that
\begin{equation}
\begin{split}
 &{}_2\varphi_1\left(\begin{matrix}{a_2}^2,0\\-p\end{matrix};p,pa_2^{-1}t\right)
 -\cfrac{({a_2}^{-2};p)_\infty e^{\pi i\log (p{a_2}^{-1})/\log p}}{(-{a_2}^2;p)_\infty(-1)^{\log t/\log p}}~
 {}_2\varphi_1\left(\begin{matrix}-{a_2}^2,0\\-p\end{matrix};p,p{a_2}^{-1}t\right)\\
 &\quad=\cfrac{(-1)^{3\alpha/2}2^{-\alpha/2} e^{(\pi i\log2+\pi^2/4)/\epsilon^2-\pi s/\epsilon}}{\pi^{1/2}\epsilon^{1-\alpha}}
  \int_L e^{-u^2/2-2^{1/2}su}u^{\alpha-1}[1+O(\epsilon)]~du,
\end{split} 
\end{equation}
as 
\begin{align}
 &\epsilon\to+0\quad\text{\rm (when ${\rm Im}(s)<0$)},\\
 &\epsilon\to-0\quad\text{\rm (when ${\rm Im}(s)>0$)}.
\end{align}
The path $L$ runs from 
$-\infty$ to $+\infty$ when ${\rm Im}(s)<0$ and $+\infty$ to $-\infty$ when ${\rm Im}(s)>0$
so that $u=0$ lies to the right of the path.
\end{lemma}
\begin{lemma}\label{lemma:continuous_weber_2}
With the substitutions \eqref{eqn:condition_limit_a4}, it follows that
\begin{equation}
\begin{split}
 &{}_2\varphi_1\left(\begin{matrix}-{a_2}^2,0\\-p\end{matrix};p,p{a_2}^{-1}t\right)\notag\\
 &\quad-\cfrac{(-{a_2}^2;p)_\infty e^{(2\log (-pa_2)\log a_2-\log (p{a_2}^{-1})\log (-{a_2}^{-2}))/\log p}}
    {{a_2}^4({a_2}^2;p)_\infty}~(-1)^{\log t/\log p}
  {}_2\varphi_1\left(\begin{matrix}{a_2}^2,0\\-p\end{matrix};p,p{a_2}^{-1}t\right)\notag\\
 &=(-1)^{-1-4\alpha}2^{-1/2+\alpha/2}\pi^{-1/2}\epsilon^{-\alpha} e^{s^2+2\pi i\log2/\epsilon^2}
   \int_{L}  e^{-u^2/2-2^{1/2}isu}u^{-\alpha}[1+O(\epsilon)]du,
\end{split} 
\end{equation}
as
\begin{align}
 &\epsilon\to+0\quad\text{\rm (when ${\rm Re}(s)<0$)},\\
 &\epsilon\to-0\quad\text{\rm (when ${\rm Re}(s)>0$)}.
\end{align}
The path $L$ runs from 
$-\infty$ to $+\infty$ when ${\rm Re}(s)>0$ and $+\infty$ to $-\infty$ when ${\rm Re}(s)<0$
so that $u=0$ lies to the right of the path.
\end{lemma}

The proofs of Lemma \ref{lemma:continuous_weber_1} and Lemma \ref{lemma:continuous_weber_2} will be given in the next section.
For these proofs, we use the saddle point method\cite{HKW:A1A1,Prellberg_1995,Wong:book}.
Moreover, for the asymptotic expansions of the $q$-shifted factorials as $q\to1^{-}$ $(q=1-\epsilon,~\epsilon>0,~ \epsilon \to 0)$, 
we use the following proposition:
\begin{proposition}[\cite{Mcintosh_1999,Prellberg_1995}]\label{prop:qshifted_asymptotic}
As $q\to1^{-}$, the $q$-shifted factorials have an asymptotic expansions,
\begin{align}
 &\log(t;q)_\infty=\cfrac{Li_2(t)}{\log q}+\cfrac{\log(1-t)}{2}+O(\log q),\\
 &\log(q;q)_\infty
 =\cfrac{\pi^2}{6\log q}+\cfrac{1}{2}\log\left(-\frac{2\pi}{\log q}\right)+O(\log q),\\
 &\log(-q;q)_\infty=-\cfrac{\pi^2}{12\log q}-\cfrac{\log2}{2}+O(\log q).
\end{align}
This is uniform for $t\in\mathbb{C}$ such that $|\arg(1-t)|<\pi$.
Here $Li_2(t)$ is the Euler dilogarithm defined by
\begin{align}
 Li_2(t)&=-\int^t_0\cfrac{\log(1-u)}{u}du\label{equation:dilog_log}\\
 &=\sum_{k=1}^\infty\cfrac{t^k}{k^2}\quad(|t|<1).
\end{align}
\end{proposition}

\subsection{Symmetric $q$-P($D_5^{(1)}$)}
Putting
\begin{equation}\label{eqn:continous_limit_d5}
 t^{1/2}=\cfrac{i(1-p)}{2}~s,\quad
 W=\cfrac{\epsilon s}{2}~z,\quad
 p=1+\epsilon,
\end{equation}
and taking the limit  $\epsilon\to -0$, symmetric $q$-{\rm P}$(D_5^{(1)})$ \eqref{eqn:sym_d5_2} yields the
Painlev\'e III equation\cite{RGH:dP}
\begin{equation}
 z''-\frac{(z')^2}{z}+\frac{z'}{s}+\frac{2(\nu_1-\nu_2z^2)}{s})-z^3+\frac{1}{z}=0.
\end{equation}
\begin{proposition}[\cite{KOS:dp3,KS:q2DTL}]
Under the parametrization \eqref{eqn:continous_limit_d5}, the hypergeometric solution in Proposition
\ref{prop:hyper_sym_d5} is reduced to
\begin{equation}
 z=\cfrac{G'}{G}-\cfrac{\nu_2}{s}~,
\end{equation}
\begin{equation}\label{eqn:sol_p3}
 G=A~J_{\nu_2}(s)+B~J_{-\nu_2}(s),
\end{equation}
with
\begin{equation}
 \nu_1=\nu_2+1.
\end{equation}
Here, $A$ and $B$ are constants and $J_\nu(x)$ is the Bessel function
\begin{equation}
 J_\nu(x)=\cfrac{x^\nu}{2^\nu \Gamma(\nu+1)}~
 {}_0F_1\left(\begin{matrix}-\\\nu+1\end{matrix};-\cfrac{x^2}{4}\right).
\end{equation}
\end{proposition}
We note that \eqref{eqn:riccati_d5} and \eqref{eqn:3term_d5} are reduced to the following equations
\begin{equation}
 z'+z^2+\cfrac{1+2\nu_2}{s}~z+1=0,
\end{equation}
\begin{equation}
 G''+\cfrac{G'}{s}+\left(1-\cfrac{{\nu_2}^2}{s^2}\right)G=0,
\end{equation}
respectively. 

\subsection{Symmetric $q$-P($E_6^{(1)}$)}
Putting
\begin{equation}\label{eqn:continuous_limit_e6}
\begin{split}
 &t=\cfrac{(\gamma^2\epsilon^2-16)^{1/2}}{4\epsilon}(1+\epsilon)^{-\log s/\epsilon},\quad
 b_1=-1+2^{1/2}\beta\epsilon,\quad
 b_3=1+2^{1/2}\alpha\epsilon,\\
 &b_5=\cfrac{(\gamma^2\epsilon^2-16)^{1/2}}{\gamma\epsilon-4},\quad
 X=\cfrac{z+1}{z-1},\quad
 p=1+\epsilon,
\end{split} 
\end{equation}
and taking the limit $\epsilon\to -0$, symmetric $q$-{\rm P}$(E_6^{(1)})$ \eqref{eqn:sym_e6} yields
the Painlev\'e V equation\cite{RG:coales}
\begin{equation}\label{eqn:p5}
 z''=\left(\frac{1}{2z}+\frac{1}{z-1}\right)(z')^2
 -\frac{z'}{s}
 +\frac{(z-1)^2}{s^2} \left(\alpha^{2} z-\cfrac{\beta^{2}}{z}\right)
 +\frac{\gamma}{s}z
 -\frac{2z(z+1)}{z-1}.
\end{equation}
\begin{proposition}\label{prop:p5_sol}
Under the parametrization \eqref{eqn:continuous_limit_e6}, the hypergeometric solution in
Proposition \ref{prop:hyper_sym_e6} is reduced to the following solution to \eqref{eqn:p5}:
\begin{equation}
 z=\frac{sG'-2^{1/2}\beta G}{sG'-(2^{1/2}\beta+2s)G}.
\end{equation}
\begin{equation}\label{eqn:sol_p5}
 G=A~{}_2F_0\left(\begin{matrix}2^{1/2}\beta+\cfrac{\gamma}{2},2^{1/2}\beta\\-\end{matrix};-\frac{1}{2s}\right)
 +\hat{B}e^{2s}s^{2^{-1}\gamma+2^{3/2}\beta-1}~
 {}_2F_0\left(\begin{matrix}1-2^{1/2}\beta-\cfrac{\gamma}{2},1-2^{1/2}\beta\\-\end{matrix};\frac{1}{2s}\right),
\end{equation}
with
\begin{equation}
 \gamma=2(1+2^{1/2}\alpha-2^{1/2}\beta).
\end{equation}
Here, $A$ and $\hat B$ are constants.
\end{proposition}
The hypergeometric solution in Proposition \ref{prop:hyper_sym_e6} 
is reduced to the following solution to \eqref{eqn:p5}
\begin{equation}
 sz'+\left(2^{1/2}\beta+\cfrac{\gamma}{2}-1\right)z^2
 -\left(2^{3/2}\beta+\cfrac{\gamma}{2}+2s-1\right)z+2^{1/2}\beta=0,
\end{equation}
\begin{equation}
 G''+\frac{2-\gamma-2^{5/2}\beta-4s}{2s}G'+\frac{\beta\gamma+2^{3/2}\beta^{2}}{2^{1/2}s^2}G=0,
\end{equation}
respectively. 
We note that \eqref{eqn:sol_p5} is obtained from
\eqref{eqn:sol_e6} by the following procedure.  By using Heine's
transformation\cite{Koekoek:book}
\begin{equation}\label{eqn:heine_1}
 {}_2\varphi_1\left(\begin{matrix}a,b\\c\end{matrix};q,z\right)
 =\cfrac{(abc^{-1},z;q)_\infty}{(z;q)_\infty}~
 {}_2\varphi_1\left(\begin{matrix}a^{-1}c,b^{-1}c\\c\end{matrix};q,abc^{-1}z\right),
\end{equation}
and setting
\begin{equation}
 \hat{B}=\cfrac{\Theta(t;p)\Theta(-b_1{b_5}^{-1}\epsilon \tilde{t};p)({b_1}^{-1}b_5t,pb_1{b_5}^{-1}t^{-1};p)_\infty}
 {\Theta(-t;p)\Theta({b_1}^{-1}b_5\epsilon t;p)(b_1{b_5}^{-1}\tilde{t},{b_1}^{-1}b_5t^{-1};p)_\infty}~
 B,
\end{equation}
\eqref{eqn:sol_e6} is rewritten as
\begin{align}\label{eqn:sol_e6_2}
 G=&A~{}_2\varphi_1\left(\begin{matrix}{b_1}^{-1}{b_5}^2,-{b_1}^{-1}\\-p\end{matrix}
  ;p,b_1{b_5}^{-1}\tilde{t}\right)\notag\\
 &+\hat{B}\cfrac{\Theta({b_1}^{-1}b_5\epsilon t;p)({b_1}^{-1}b_5t^{-1};p)_\infty}
  {\Theta(-b_1{b_5}^{-1}\epsilon \tilde{t};p)(pb_1{b_5}^{-1}t^{-1};p)_\infty}~
 {}_2\varphi_1\left(\begin{matrix} pb_1{b_5}^{-2},-pb_1\\ -p\end{matrix};p,{b_1}^{-1}b_5t\right).
\end{align}
Then \eqref{eqn:sol_e6_2} yields the hypergeometric series in
\eqref{eqn:sol_p5} by taking a term-by-term limit. We use Proposition
\ref{prop:qshifted_asymptotic} for the limit of the coefficient.
\begin{remark}
Setting
\begin{equation}
 s=-\frac{t}{2},\quad z(s)=1-\frac{1}{f(t)},\quad G(s)=e^{2s}s^{a_2}\phi(t),\quad \alpha=-2^{-1/2}a_1,\quad\beta=2^{-1/2}a_2,
\end{equation}
we can rewrite the Riccati equation, the dependent variable teansformation and the linear differential equation as
\begin{align}
 &tf'-tf(1-f)+(a_1+a_2)f-a_1=0,\\
 &f=\cfrac{\phi'}{\phi},\\
 &t\phi''+(a_1+a_2-t)\phi'-a_1\phi=0,\label{eqn:linear_p5}
\end{align}
respectively.
These coincide with the result in \cite{Masuda:p5,Okamoto:p5}.
We note here that in \cite{Masuda:p5,Okamoto:p5} the solution to \eqref{eqn:linear_p5} is given by the hypergeometric series around zero 
while in this section that is given by the series at infinity.
\end{remark}
\subsection{Symmetric $q$-P($E_7^{(1)}$)}
Putting
\begin{equation}\label{eqn:continuous_limit_e7}
 b_1=e^{(\beta+1/2)\epsilon},\quad 
 b_3=-e^{(\alpha+1/2)\epsilon},\quad 
 b_5=e^{\delta\epsilon},\quad 
 b_7=-e^{\gamma\epsilon},\quad 
 p=e^{\epsilon},
\end{equation}
and taking the limit $\epsilon\to -0$, symmetric $q$-{\rm P}$(E_7^{(1)})$ \eqref{eqn:sym_e7} yields
Painlev\'e VI equation
\begin{align}\label{eqn:p6}
 X''=&\cfrac{1}{2}\left(\cfrac{1}{X+1}+\cfrac{1}{X-1}+\cfrac{1}{X+t}+\cfrac{1}{X-t}\right)\left(X'\right)^2\notag\\
 &-\left(\cfrac{1}{t}+\cfrac{1}{t-1}+\cfrac{1}{t+1}+\cfrac{1}{X-t}-\cfrac{1}{X+t}\right)X'\notag\\
 &+\cfrac{(X^2-t^2)(X^2-1)}{t^2(t^2-1)}
 \left(\cfrac{(4\alpha^2-1)t}{4(X+t)^2}-\cfrac{(4\beta^2-1)t}{4(X-t)^2}
 -\cfrac{\gamma^2}{(X+1)^2}+\cfrac{\delta^2}{(X-1)^2}\right),
\end{align}
where $X'={\rm d}X/{\rm d}t$\cite{GR:qP6}. 
\begin{proposition}
Under the parametrization \eqref{eqn:continuous_limit_e7}, the hypergeometric solution in
Proposition \ref{prop:hyper_sym_e7} is reduced to the following solution to \eqref{eqn:p6}:
\begin{equation}
 X(t)=\frac{(t-1)t F'-(1+\alpha+\beta+t(\alpha+\beta+2\delta+1))F}
 {(t-1)t F'-(1+\alpha+\beta+2\delta+t(\alpha+\beta+1))F}.
\end{equation}
\begin{align}\label{eqn:sol_p6}
 F=&A~{}_2F_1\left(\begin{matrix}\alpha+1,\alpha+\beta+\delta+1\\\alpha+\delta+1\end{matrix}
  ;\left(\cfrac{1+t}{1-t}\right)^2\right)\notag\\
 &+\hat{B}~\cfrac{(1-t)^{2(1+2\alpha+\beta+\delta)}}{(1+t)^{2(\alpha+\beta)}t^{1+\alpha+\beta}}~
 {}_2F_1\left(\begin{matrix}-\alpha-\beta-\delta,-\alpha\\-\alpha-\beta\end{matrix}
  ;-\cfrac{4t}{(1-t)^2}\right),
\end{align}
with 
\begin{equation}
 \alpha+\beta+\delta+\gamma+1=0.
\end{equation}
Here, $A$ and $\hat B$ are constants.
\end{proposition}
We note that \eqref{eqn:riccati_e7} and \eqref{eqn:3term_2_e7} are reduced to the following equations
\begin{equation}
 X'=\frac{1+\alpha+\beta+2\delta-t(\alpha-\beta)}{t(t^2-1)}X^2 
 +\frac{\alpha+\beta+1}{t}X
 +\frac{\alpha-\beta-t(\alpha+\beta+2\delta+1)}{t^2-1},
\end{equation}
\begin{equation}
 F''-\frac{2+\alpha+\beta+2(2+3\alpha+\beta+2\delta)t+(\alpha+\beta)t^2}{t(t^2-1)}F'
 +\frac{4(\alpha+1)(\alpha+\beta+\delta+1)}{t(t-1)^2}F=0,
\end{equation}
respectively. We also note that \eqref{eqn:sol_p6} is obtained by the following procedure.  By using
the transformation (see, (III.23) in \cite{Gasper-Rahman:BHS})
\begin{equation}
 {}_8W_7\left(a;b,c,d,e,f;q,\cfrac{q^2a^2}{bcdef}\right)
 =\cfrac{\left(qa,\cfrac{qa}{ef},\cfrac{q^2a^2}{bcde},\cfrac{q^2a^2}{bcdf};q\right)_\infty}
 {\left(\cfrac{qa}{e},\cfrac{qa}{f},\cfrac{q^2a^2}{bcd},\cfrac{q^2a^2}{bcdef};q\right)_\infty}~
 {}_8W_7\left(\cfrac{qa^2}{bcd};\cfrac{qa}{cd},\cfrac{qa}{bd},\cfrac{qa}{bc},e,f;q,\cfrac{qa}{ef}\right),
\end{equation}
and setting
\begin{equation}
 \hat{B}=B\cfrac{\Theta(t;p)}{\Theta(-b_1b_3t;p)(-b_1b_3)^{\log t/\log p}}\, ,
\end{equation}
\eqref{eqn:sol_e7} is rewritten as
\begin{equation}\label{eqn:sol_e7_2}
\begin{split}
 &F=AF_1+\hat{B}\hat{F}_2,\\
 &F_1=
 {}_8W_7\left(-{b_3}^2 b_5;p^{1/2}b_3,-p^{1/2}b_3,-b_1 b_3 b_5,
 b_3b_5\tilde{t},b_3 b_5\tilde{t}^{-1};p,p{b_1}^{-1}{b_3}^{-1}{b_5}^{-1}\right),\\
 &\hat{F}_2=\cfrac{(-{b_3}^{-1}\tilde{t},
 -p^{1/2}{b_1}^{-1}{b_3}^{-1}{b_5}^{-1}\tilde{\tilde{t}},
 p^{1/2}{b_1}^{-1}{b_3}^{-1}{b_5}^{-1}\tilde{\tilde{t}},
 {b_3}^{-1}\tilde{\tilde{t}}, 
 b_3b_5\tilde{t}, 
 -p{b_1}^{-1}{b_3}^{-1}, 
 -{b_3}^{-1}\tilde{\tilde{t}};p)_{\infty }}
 {(b_1 b_5 \tilde{t},
 -b_3 \tilde{\tilde{t}},
 p^{1/2}\tilde{t},
 -p^{1/2}\tilde{t},
 {b_3}^{-1}{b_5}^{-1}
 \tilde{\tilde{t}},
 -p^3{b_1}^{-1}{b_3}^{-2}{b_5}^{-1}t,
 -b_5;p)_{\infty }}\\
 &~\times\cfrac{\Theta(-b_1b_3t;p)}{\Theta(t;p)}\\
 &~\times
 {}_8W_7\left(-{b_1}^{-1}{b_3}^{-2}{b_5}^{-1}\tilde{\tilde{t}};
  -p{b_1}^{-1}{b_3}^{-1}{b_5}^{-1},  -p^{1/2}{b_3}^{-1},
  p^{1/2}{b_3}^{-1}, {b_3}^{-1}{b_5}^{-1}\tilde{\tilde{t}},
  p{b_1}^{-1}{b_3}^{-1}{b_5}^{-1};
  p,b_3b_5\tilde{t}\right).
\end{split}
\end{equation}
Then \eqref{eqn:sol_p6} is obtained from \eqref{eqn:sol_e7_2} by taking a
term-by-term limit and using Proposition \ref{prop:qshifted_asymptotic}.
\section{Proof of Lemma \ref{lemma:continuous_weber_1} and \ref{lemma:continuous_weber_2}}
In this section, we give the proof of Lemma
\ref{lemma:continuous_weber_1} and \ref{lemma:continuous_weber_2}.
Taking the continuous limit term-by-term in the hypergeometric series in
\eqref{eqn:sol_a4} does not yield any meaningful result.  In this case,
we use the saddle point method\cite{HKW:A1A1,Prellberg_1995,Wong:book}.
Consider the complex integral
\begin{equation}
 I(s)=\int_C e^{-g(z)/\epsilon^2}f(z)~dz,
\end{equation}
where $f(z)$ and $g(z)$ are analytic functions.
The point $z=z_0$ satisfying $g'(z_0)=0$ and the direction
\begin{equation}
 \arg(z-z_0)=\cfrac{\pi}{2}-\cfrac{1}{2}\arg\left(-\cfrac{g''(z_0)}{\epsilon^2}\right),
\end{equation}
are called a saddle point and a steepest descent direction,
respectively.  When the path $C$ passes in the steepest descent
direction, $|e^{-g(z)/\epsilon^2}f(z)|$ reaches its peak at the saddle
point.  Therefore $I(s)$ may be evaluated only around the saddle point when
$\epsilon\to0$.  For the asymptotic expansions of the $q$-shifted
factorials as $q\to1^{-}$, we use Proposition \ref{prop:qshifted_asymptotic}.

We here fix the branch of $\log$ and fractional power functions as
\begin{align}
 &\log z=\log |z|+i\arg z\quad (0\leq\arg z<2\pi),\\
 &z^{m/n}=e^{(m/n)\log z}=|z|^{m/n}e^{(m/n)i\arg z}\quad
 (m\in\mathbb{Z},~n\in\mathbb{Z}_{>0},~0\leq\arg z<2\pi),
\end{align}
where $z\in\mathbb{C^\ast}$.
Note that $\log(XY)=\log X+\log Y$ and $\log(X/Y)=\log X-\log Y$
are valid only $\mod 2\pi i$.
\subsection{Proof of Lemma \ref{lemma:continuous_weber_1}}
\begin{figure}[h]
\begin{minipage}{0.45\hsize}
\begin{center}
\includegraphics[width=1\textwidth]{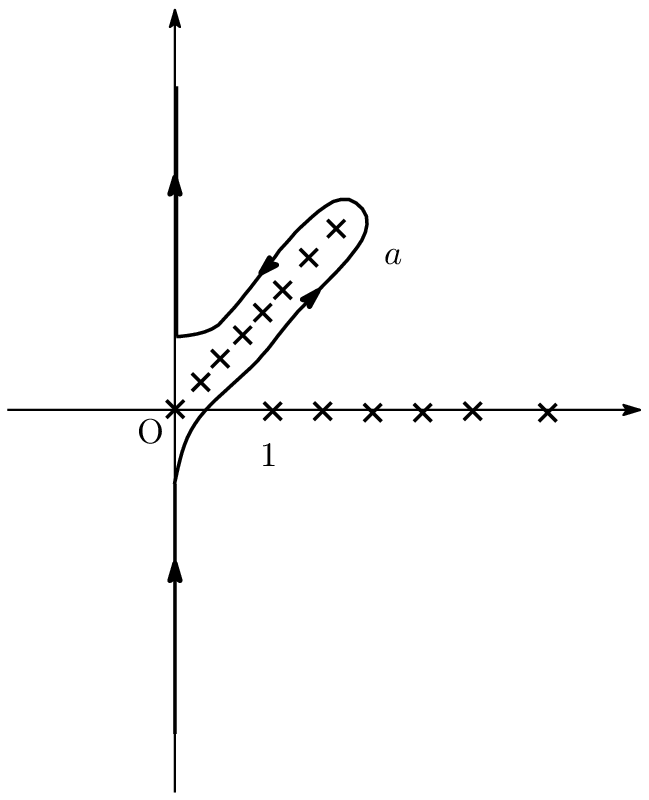}
\end{center}
\caption{Path of integration $C$}
\label{fig:contour_C}
\end{minipage}
\quad
\begin{minipage}{0.45\hsize}
\begin{center}
\includegraphics[width=1\textwidth]{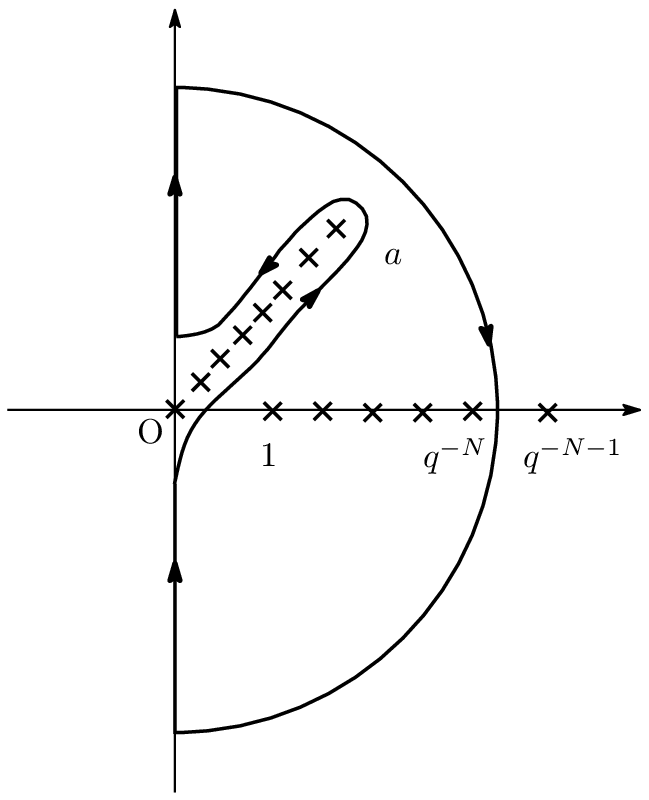}
\end{center}
\caption{contour of integration $C_N^0$}
\label{fig:contour_CN}
\end{minipage}
\end{figure}

We consider the continuous limit of
\begin{equation}\label{eqn:def_A1}
 A_1={}_2\varphi_1\left(\begin{matrix}{a_2}^2,0\\-p\end{matrix};p,pa_2^{-1}t\right),
\end{equation}
with
\begin{equation}\label{eqn:condition_limit_a4_series}
 t=2ie^{is\epsilon-3\alpha\epsilon^2/2},\quad
 a_2=ie^{-\alpha\epsilon^2/2},\quad
 p=e^{-\epsilon^2},\quad
 \epsilon\to\pm0.
\end{equation}
We note that in taking the limit, the sign of $\epsilon$ is chosen according to the value of
$s$ as shown later.  By using Heine's transformation\cite{Koekoek:book}
\begin{equation}\label{eqn:heine_2}
 {}_2\varphi_1\left(\begin{matrix}a,0\\c\end{matrix};q,z\right)
 =\cfrac{(az;q)_\infty}{(c,z;q)_\infty}~
 {}_1\varphi_1\left(\begin{matrix}z\\az\end{matrix};q,c\right),
\end{equation}
$A_1$ is rewritten as
\begin{equation}
 A_1=\cfrac{(pa_2t;p)_\infty}{(-p,p{a_2}^{-1}t;p)_\infty}~
 {}_1\varphi_1\left(\begin{matrix}p{a_2}^{-1}t\\pa_2t\end{matrix};p,-p\right).
\end{equation}
We next prepare a suitable integral representation for $A_1$.
\begin{lemma}\label{lemma:1phi1_integral}
It holds that
\begin{equation}\label{eqn:1phi1_integral}
 {}_1\varphi_1\left(\begin{matrix}a\\c\end{matrix};q,x\right)
 =\cfrac{(a;q)_\infty(q;q)_\infty}{2\pi i(c;q)_\infty}
  \int_C\cfrac{z^{-\log x/\log q}(cz^{-1};q)_\infty}{(az^{-1};q)_\infty(z;q)_\infty}~dz,
\end{equation}
where the path $C$ runs from $-i\infty$ to $i\infty$
so that the poles of $1/(z;q)_\infty$ lie to the right of the path
and the other poles lie to the left of the path $($as shown in Figure \ref{fig:contour_C}$)$.
\end{lemma}
{\bf Proof.} 
We derive \eqref{eqn:1phi1_integral} according to the method in \cite{Gasper-Rahman:BHS,Prellberg_1995,Watson:1910}.
We consider the integral
\begin{equation}
 \oint_{C_N^0}\cfrac{z^{-\log x/\log q}(cz^{-1};q)_\infty}{(az^{-1};q)_\infty(z;q)_\infty}~dz,
\end{equation}
where the contour $C_N^0$ 
is a large clockwise-oriented semicircle of radius $q^{-N-1/2}$ with center at the origin
and circle round only part of the poles of $1/(z;q)_\infty$ (as shown in Figure \ref{fig:contour_CN}).
By the residue theorem, we have
\begin{equation}
 \cfrac{1}{2\pi i}
 \oint_{C_N^0}\cfrac{z^{-\log x/\log q}(cz^{-1};q)_\infty}{(az^{-1};q)_\infty(z;q)_\infty}~dz
 =\cfrac{(c;q)_\infty}{(a;q)_\infty(q;q)_\infty}
  \sum_{n=0}^N 
  \cfrac{(a;q)_n}{(c;q)_n(q;q)_n}(-1)^nq^{n(n-1)/2}x^n.
\end{equation}
Therefore we obtain
\begin{equation}\label{eqn:1phi1_ointegral}
 {}_1\varphi_1\left(\begin{matrix}a\\c\end{matrix};q,x\right)
 =\cfrac{(a;q)_\infty(q;q)_\infty}{2\pi i(c;q)_\infty}
  \lim_{N\to\infty}
  \oint_{C_N^0}\cfrac{z^{-\log x/\log q}(cz^{-1};q)_\infty}{(az^{-1};q)_\infty(z;q)_\infty}~dz.
\end{equation}
In order to estimate the contribution from
\begin{equation}
 C^1_N=\left\{z=q^{-N-1/2}e^{it}:|t|<\frac{\pi}{2}\right\},
\end{equation}
we need a bound on the integrand for large $|z|$
\begin{equation}
 \sup_{|z|=q^{-N-1/2}}\left|\cfrac{z^{-\log x/\log q}}{(z;q)_\infty}\right|
 \leq\left|\cfrac{|x|^{N+1/2}e^{-\pi^2/\log q}}{(q^{-1/2};q)_\infty\prod_{n=1}^N(1-q^{-1/2-n})}\right|
 =O(q^{N^2/2}),
\end{equation}
as it is dominated by the product in the denominator.
From
\begin{equation}
 \left|\int_{C^1_N}\cfrac{z^{-\log x/\log q}(cz^{-1};q)_\infty}{(az^{-1};q)_\infty(z;q)_\infty}dz\right|
 \leq \pi q^{-N-1/2}
 \sup_{|z|=q^{-N-1/2}}
 \left|\cfrac{z^{-\log x/\log q}(cz^{-1};q)_\infty}{(az^{-1};q)_\infty(z;q)_\infty}\right|
 =O(q^{N^2/2}),
\end{equation}
and \eqref{eqn:1phi1_ointegral}, we have completed the proof.\hfill\qed

\begin{figure}[h]
\begin{center}
\includegraphics[width=0.5\textwidth]{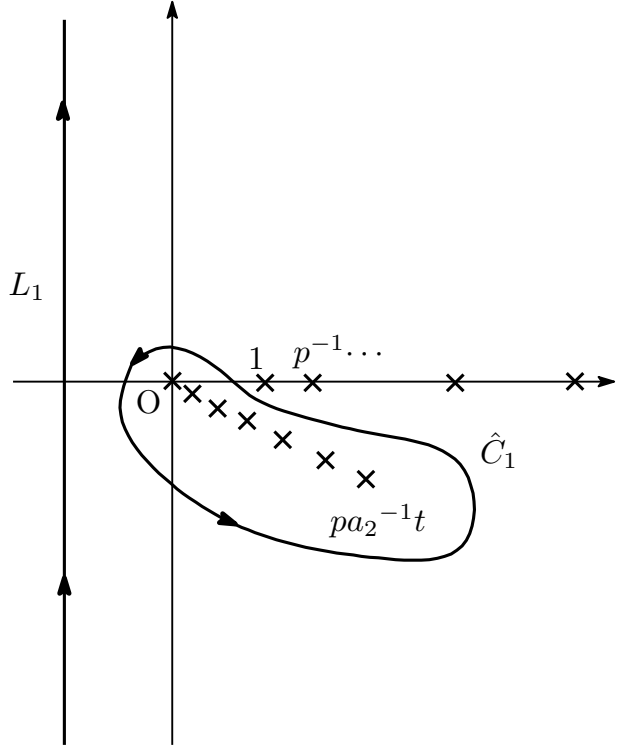}
\end{center}
\caption{The paths of integration $L_1$ and $\hat{C}_1$}
\label{fig:contour_L1_hatC1}
\end{figure}

By using Lemma \ref{lemma:1phi1_integral},
it holds that
\begin{equation}\label{eqn:A1_C1}
 A_1=\cfrac{(p;p)_\infty}{2\pi i(-p;p)_\infty}~
 \int_{C_1}\cfrac{(pa_2tz^{-1};p)_\infty e^{\pi i\log z/\log p}}{z(p{a_2}^{-1}tz^{-1};p)_\infty(z;p)_\infty}~dz,
\end{equation}
where the path $C_1$ runs from $-i\infty$ to $i\infty$
so that the poles of $1/(z;p)_\infty$ lie to the right of the path
and the other poles lie to the left of the path.
We divide the integral path $C_1$ into $L_1$ which runs from $-i\infty$ to $i\infty$
so that all poles lie to the right of the path
and $\hat{C}_1$ which is an anticlockwise-oriented contour and encircles
all poles except for the poles of $1/(z;p)_\infty$ (as shown in Figure \ref{fig:contour_L1_hatC1}).

We now evaluate $A_1$ at $\epsilon\to 0$ by using the saddle point method.

{\boldmath\bf (1) Path $L_1$.}\quad
Set
\begin{equation}
 I_1(s)=\int_{L_1}\cfrac{(pa_2tz^{-1};p)_\infty e^{\pi i\log z/\log p}}{z(p{a_2}^{-1}tz^{-1};p)_\infty(z;p)_\infty}~dz.
\end{equation}
Using Proposition \ref{prop:qshifted_asymptotic}, we rewrite $I_1(s)$ as
\begin{equation}\label{eqn:I1_fg}
 I_1(s)=\int_{L_1}e^{-g_1(z)/\epsilon^2}f_1(z)[1+O(\epsilon)]~dz,
\end{equation}
where
\begin{align}
 &g_1(z)=Li_2\left(-\cfrac{2}{z}\right)-Li_2\left(\cfrac{2}{z}\right)-Li_2(z)
  +\pi i\log{z}+is\epsilon\log\left(\cfrac{z-2}{z+2}\right),\\
 &f_1(z)=e^{-2s^2z/(z^2-4)}(z-2)^{1/2+\alpha}(z+2)^{-1/2-2\alpha}(1-z)^{-1/2}z^{\alpha-1}.\label{eqn:f1}
\end{align}
By the identity of dilogarithm\cite{AN_Kirillov:dilogarithm}
\begin{equation}
 Li_2(z^{-1})=-Li_2(z)-\cfrac{(\log z)^2}{2}-\pi i\log z+\cfrac{\pi^2}{3},
\end{equation}
$g_1(z)$ is rewritten as
\begin{align}
 g_1(z)&=Li_2\left(\cfrac{z}{2}\right)-Li_2\left(-\cfrac{z}{2}\right)-Li_2(z)
  +\pi i\log2+\cfrac{3\pi^2}{2}+is\epsilon\log\left(\cfrac{z-2}{z+2}\right)\label{eqn:g_1}\\
 &=\pi i\log2+\cfrac{3\pi^2}{2}+\pi s\epsilon-is\epsilon z-\cfrac{z^2}{4}+O(z^3)\quad (|z|<1).\label{eqn:g1_z}
\end{align}
Let us first find a zero point of $g_1'(z)$ (saddle point).
Differentiating \eqref{eqn:g_1} by $z$ and using \eqref{equation:dilog_log},
we obtain
\begin{align}
 g_1'(z)&=\cfrac{1}{z}\log\left(1+\cfrac{z^2}{z-2}\right)+\cfrac{4is\epsilon}{z^2-4}\\
 &=-is\epsilon-\cfrac{z}{2}-\cfrac{1+is\epsilon}{4}~z^2+O(z^3)\quad (|z|<1).
 \label{eqn:g_1'(z)_expansion}
\end{align}
We note here that only 
$z=0,~z=\pm 2,~z=\infty$ can be saddle points at $\epsilon\to0$.
\begin{lemma}\label{lemma:Rouche}
There exists only one saddle point in the domain $|z|\leq1/2$.
\end{lemma}
\noindent{\bf Proof.}
Set
\begin{equation}
 A(z)=-\cfrac{4is\epsilon}{z^2-4}.
\end{equation}
Functions $g_1'(z)$ and $A(z)$ are regular in $|z|\leq 1/2$.
Furthermore, since 
\begin{equation}
 |A(z)|<<1,
\end{equation}
it is obvious that
\begin{equation}
 |g_1'(z)|>|A(z)|,
\end{equation}
on the circle $|z|=1/2$.
By Rouch\'e's theorem, the numbers of zeros of $A(z)+g_1'(z)=\log\left(1+\frac{z^2}{z-2}\right)/z$ are equal to those of $g_1'(z)$. 
Therefore we have completed the proof. \hfill\qed
\par \medskip

From Lemma \ref{lemma:Rouche}, the saddle point $z_1$ is determined by assuming the series expansion 
$z=\sum_{k=1}^\infty C_k\epsilon^k$ in \eqref{eqn:g_1'(z)_expansion} as
\begin{equation}
 z_1=-2is\epsilon+2s^2\epsilon^2+O(\epsilon^3).
\end{equation}
Since
\begin{equation}\label{eqn:g1''z1}
 g_1''(z_1)=-\cfrac{1}{2}+O(\epsilon),
\end{equation}
the steepest descent direction is given as
\begin{equation}
 \arg(z-z_1)=\cfrac{\pi}{2}-\cfrac{1}{2}\arg\left(\cfrac{1+O(\epsilon)}{2\epsilon^2}\right)\sim \cfrac{\pi}{2}.
\end{equation}
In order that $L_1$ passes in the steepest descent direction, 
it is necessary from the configuration of the paths (see Figure \ref{fig:contour_L1_hatC1}) that
\begin{equation}
 {\rm Re}(z_1)=2{\rm Im}(s)\epsilon+2\left({{\rm Re}(s)}^2-{{\rm Im}(s)}^2\right)\epsilon^2+O(\epsilon^3)<0.
\end{equation}
Therefore we obtain the following condition
\begin{align}
 &\epsilon>0\quad\text{\rm (when ${\rm Im}(s)<0$)},\\
 &\epsilon<0\quad\text{\rm (when ${\rm Im}(s)>0$)}.
\end{align}
Now, we consider an approximation of the integrand of $I_1(s)$.
Since the integrand of $I_1(s)$ is evaluated only around the saddle point $z=z_1$,
we assume that $z$ is in the set
\begin{equation}
 D=\left\{\left.z=\epsilon^k A(\epsilon)~\right|~k\in\mathbb{R}_{>0},~0<|A(0)|<\infty\text{ or }A(\epsilon)=0\right\}.
\end{equation}
For simplicity, we first change the variable $z$ to $\hat{z}$ so that the saddle point is given as $\hat{z}=-2is\epsilon$.
From \eqref{eqn:g1''z1}, we set $\hat{z}$ so as to satify
\begin{equation}\label{eqn:g_1'_hatz}
 g_1'(z)dz=-\cfrac{1}{2}~(\hat{z}+2is\epsilon) d\hat{z}.
\end{equation}
Moreover, from \eqref{eqn:g1_z} and to calculate the integration by substitution from $z$ to $\hat{z}$ simply, 
we also require
\begin{equation}\label{eqn:g1_hatz}
 g_1(z)=-\cfrac{\hat{z}^2}{4}-is\epsilon \hat{z} +\pi i\log2+\cfrac{3\pi^2}{2}+\pi s\epsilon+2i s\epsilon^3+\epsilon^4,
\end{equation}
that is,
\begin{equation}\label{eqn:relation_z_hatz_1}
 \hat{z}^2+4is\epsilon \hat{z}-8i s\epsilon^3-4\epsilon^4=4is\epsilon z+z^2+O(z^3).
\end{equation}
We now add a condition between $z$ and $\hat{z}$ so that the correspondence $z\leftrightarrow\hat{z}$ is $1$ : $1$.
To this end, we fix the branch at the point $\hat z=\hat{w}_0$ corresponding to $z=0$.
From \eqref{eqn:relation_z_hatz_1} ${\hat{w}_0}$ satisfies
\begin{equation}
 {\hat{w}_0}^2+4is\epsilon \hat{w}_0-8i s\epsilon^3-4\epsilon^4
 = ({\hat{w}_0}-2\epsilon^2)({\hat{w}_0}+4is\epsilon+2\epsilon^2)
 =0.
\end{equation}
We fix the branch by choosing $\hat{w}_0=2\epsilon^2$, that is, from \eqref{eqn:relation_z_hatz_1} $\hat{z}$ is expressed by $z$ as
\begin{equation}\label{eqn:relation_z_hatz_2}
 \hat{z}=-2is\epsilon+2is\epsilon\sqrt{1-\frac{2i\epsilon}{s}-\frac{\epsilon^2}{s^2}-\frac{z^2+4is\epsilon z+O(z^3)}{4s^2\epsilon^2}}~.
\end{equation}
Furthermore, $z$ can be also expressed by $\hat{z}$.
Noticing that $z=\epsilon^k A(\epsilon)\in D$, we obtain from \eqref{eqn:relation_z_hatz_1}
\begin{equation}
 z=-2is\epsilon\pm 2is\epsilon\sqrt{1+\frac{2i\epsilon}{s}+\frac{\epsilon^2}{s^2}-\frac{\hat{z}^2+4is\epsilon \hat{z}+O\left(A(\epsilon)^3\epsilon^{3k}\right)}{4s^2\epsilon^2}}~.
\end{equation}
Since $z=0$ corresponds to $\hat{z}=\hat{w}_0$, it holds that
\begin{equation}\label{eqn:relation_z_hatz_3}
 z=-2is\epsilon+2is\epsilon\sqrt{1+\frac{2i\epsilon}{s}+\frac{\epsilon^2}{s^2}-\frac{\hat{z}^2+4is\epsilon \hat{z}+O\left(A(\epsilon)^3\epsilon^{3k}\right)}{4s^2\epsilon^2}}~.
\end{equation}
For later convenience, we consider the point $z=w_0$ corresponding to $\hat{z}=0$.
By the assumption, $w_0$ is given in the form
\begin{equation}
 w_0=\epsilon^{k_0}A_0(\epsilon),
\end{equation}
where $k_0\in\mathbb{R}_{>0}$ and $0<|A_0(0)|<\infty$.
From \eqref{eqn:relation_z_hatz_1}, we have
\begin{equation}\label{eqn:A0_k0}
 \epsilon^{2k_0}{A_0}^2+4is\epsilon^{k_0+1}A_0+8is\epsilon^3+4\epsilon^4=O(\epsilon^{3k_0}).
\end{equation}
Comparing the degree of $\epsilon$ of \eqref{eqn:A0_k0}, we obtain $k_0=1$ or $k_0=2$.
From \eqref{eqn:relation_z_hatz_3}, it holds that
\begin{equation}
 w_0=-2is\epsilon+2is\epsilon\sqrt{1+O(\epsilon)}=O(\epsilon^2)~,
\end{equation}
which indicates
\begin{equation}
 k_0=2.
\end{equation}
Therefore we can determine $w_0$ from \eqref{eqn:A0_k0} as 
\begin{equation}
 w_0=-2\epsilon^2+O(\epsilon^3).
\end{equation}
We proceed to the approximation of integrand of $I_1(s)$ by changing the variable from $z$ to $\hat{z}$.
We expand $f_1(z)$ in the form
\begin{equation}\label{eqn:expansion_f_1}
 f_1(z)\cfrac{dz}{d\hat{z}}=\sum_{k=0}^\infty C_k(\hat{z}-\hat{w}_0)^{k+\rho}.
\end{equation}
Since the characteristic exponent of $f_1(z)$ at $z=0~(\hat{z}=\hat{w}_0=2\epsilon^2)$ is $\alpha-1$ (see \eqref{eqn:f1}), it holds that
\begin{equation}
 \rho=\alpha-1.
\end{equation}
From \eqref{eqn:I1_fg}, \eqref{eqn:g1_hatz} and \eqref{eqn:expansion_f_1}, we obtain
\begin{align}
 I_1(s)
 =&e^{(\pi i\log2+\pi^2/2)/\epsilon^2-\pi s/\epsilon}
  \int_{L_1}\sum_{k=0}^\infty C_ke^{(\hat{z}^2+4is\epsilon \hat{z})/4\epsilon^2}
  (\hat{z}-2\epsilon^2)^{k+\alpha-1}[1+O(\epsilon)]d\hat{z}\\
 =&e^{(\pi i\log2+\pi^2/2)/\epsilon^2-\pi s/\epsilon}C_0
  \int_{L_1}e^{(\hat{z}^2+4is\epsilon \hat{z})/4\epsilon^2}
  \hat{z}^{\alpha-1}[1+O(\epsilon)]d\hat{z}.\label{eqn:I1_c0}
\end{align}
In the derivation of \eqref{eqn:I1_c0}, we note that since $f_1(z)$ does not diverge at $\epsilon\to0$, $C_k$ also does not diverge.
$C_0$ in \eqref{eqn:I1_c0} can be determined in the following manner.
From \eqref{eqn:g_1'_hatz} and \eqref{eqn:expansion_f_1}, we heve
\begin{equation}\label{eqn:determine_C0}
 -\cfrac{f_1(z)(\hat{z}+2is\epsilon)}{2g_1'(z)}=\sum_{k=0}^\infty C_k(\hat{z}-2\epsilon^2)^{k+\alpha-1}.
\end{equation}
Substituting $z=w_0~(\hat{z}=0)$ in \eqref{eqn:determine_C0}, we get
\begin{equation}
 C_0=(-1)^{\alpha+1/2}2^{-\alpha}.
\end{equation}
In order to adjust to the integral representation of the Weber function, 
we introduce the variable $u$ by $\hat{z}=2^{1/2}i\epsilon u$. 
Then we obtain 
\begin{align}
 I_1(s)
 =&(-1)^{\alpha/2}2^{\alpha/2}\epsilon^\alpha e^{(\pi i\log2+\pi^2/2)/\epsilon^2-\pi s/\epsilon}
  C_0\int_{\hat{L}_1}e^{-u^2/2-2^{1/2}su}u^{\alpha-1}[1+O(\epsilon)] du\\
 =&(-1)^{3\alpha/2+1/2}2^{-\alpha/2}\epsilon^\alpha e^{(\pi i\log2+\pi^2/2)/\epsilon^2-\pi s/\epsilon}
  \int_{\hat{L}_1}e^{-u^2/2-2^{1/2}su}u^{\alpha-1}[1+O(\epsilon)] du,\label{eqn:value_I1}
\end{align}
where the path of integration $\hat{L}_1$ runs from 
$-\infty$ to $\infty$ when ${\rm Im}(s)<0$ and $\infty$ to $-\infty$ when ${\rm Im}(s)>0$
so that $u=0$ lies to the right of the path.

{\boldmath\bf (2) Contour $\hat{C}_1$.}\quad
Set
\begin{equation}
 I_2(s)=\int_{\hat{C}_1}\cfrac{(pa_2tz^{-1};p)_\infty e^{\pi i\log z/\log p}}{z(p{a_2}^{-1}tz^{-1};p)_\infty(z;p)_\infty}~dz.
\end{equation}
To evaluate $I_2(s)$, we prepare the following lemma:
\begin{lemma}\label{lemma:2phi1_integral_1}
It holds that
\begin{equation}\label{eqn:1phi1_integral2}
 \int_C\cfrac{(cz^{-1};q)_\infty e^{\pi i\log z/\log q}}{z(az^{-1};q)_\infty(z;q)_\infty}~dz
 =\cfrac{2\pi i(ac^{-1};q)_\infty e^{\pi i\log a/\log q}}{(q;q)_\infty(a;q)_\infty}~
 {}_2\varphi_1\left(\begin{matrix}a,qac^{-1}\\0\end{matrix};q,-\cfrac{c}{a}\right),
\end{equation}
where $C$ is an anticlockwise-oriented contour and encircles 
all poles except for the poles of $1/(z;q)_\infty$.
\end{lemma}
{\bf Proof.} We consider the integral
\begin{equation}
 \oint_{C_N}\cfrac{(cz^{-1};q)_\infty e^{\pi i\log z/\log q}}{z(az^{-1};q)_\infty(z;q)_\infty}~dz,
\end{equation}
where $C_N$ is an anticlockwise-oriented contour and encircles the poles of $1/(az^{-1};q)_N$.
By the residue theorem, we obtain
\begin{equation}
 \cfrac{1}{2\pi i}
 \oint_{C_N}\cfrac{(cz^{-1};q)_\infty e^{\pi i\log z/\log q}}{z(az^{-1};q)_\infty(z;q)_\infty}~dz
 =\cfrac{(ac^{-1};q)_\infty e^{\pi i\log a/\log q}}{(q;q)_\infty(a;q)_\infty}
 \sum_{k=0}^{N-1}
 \cfrac{(a;q)_k(qac^{-1};q)_k}{(q;q)_k}\left(-\cfrac{c}{a}\right)^k.
\end{equation}
Then the statement follows from the limit of $N\to\infty$.\hfill\qed

By using Lemma \ref{lemma:2phi1_integral_1}, $I_2(s)$ is rewritten as
\begin{equation}
 I_2(s)
 =\cfrac{2\pi i({a_2}^{-2};p)_\infty e^{\pi i\log (p{a_2}^{-1}t)/\log p}}{(p;p)_\infty(p{a_2}^{-1}t;p)_\infty}~
 {}_2\varphi_1\left(\begin{matrix}p{a_2}^{-1}t,p{a_2}^{-2}\\0\end{matrix};p,-{a_2}^2\right),
\end{equation}
Finally by using Heine's transformation\cite{Koekoek:book}
\begin{equation}\label{eqn:heine_3}
 {}_2\varphi_1\left(\begin{matrix}a,b\\0\end{matrix};q,z\right)
 =\cfrac{(az,b;q)_\infty}{(z;q)_\infty}~
 {}_2\varphi_1\left(\begin{matrix}z,0\\az\end{matrix};q,b\right),
\end{equation}
we obtain
\begin{equation}\label{eqn:value_I2}
 I_2(s)
 =\cfrac{2\pi i({a_2}^{-2},-p;p)_\infty e^{\pi i\log (p{a_2}^{-1})/\log p}}{(-{a_2}^2,p;p)_\infty(-1)^{\log t/\log p}}~
 {}_2\varphi_1\left(\begin{matrix}-{a_2}^2,0\\-p\end{matrix};p,p{a_2}^{-1}t\right).
\end{equation}
From \eqref{eqn:A1_C1}, \eqref{eqn:value_I1} and \eqref{eqn:value_I2}, we have
\begin{align}
 A_1
 =&\cfrac{(p;p)_\infty}{2\pi i(-p;p)_\infty}\left(I_1(s)+I_2(s)\right)\\
 =&\cfrac{(-1)^{3\alpha/2}2^{-\alpha/2} e^{(\pi i\log2+\pi^2/4)/\epsilon^2-\pi s/\epsilon}}{\pi^{1/2}\epsilon^{1-\alpha}}
  \int_{\hat{L}_1} e^{-u^2/2-2^{1/2}su}u^{\alpha-1}[1+O(\epsilon)]~du\notag\\
  &+\cfrac{({a_2}^{-2};p)_\infty e^{\pi i\log (p{a_2}^{-1})/\log p}}{(-{a_2}^2;p)_\infty(-1)^{\log t/\log p}}~
 {}_2\varphi_1\left(\begin{matrix}-{a_2}^2,0\\-p\end{matrix};p,p{a_2}^{-1}t\right),
\end{align}
which proves Lemma \ref{lemma:continuous_weber_1}.
\subsection{Proof of Lemma \ref{lemma:continuous_weber_2}}
We consider the continuous limit of
\begin{equation}
 A_2={}_2\varphi_1\left(\begin{matrix}-{a_2}^2,0\\-p\end{matrix};p,p{a_2}^{-1}t\right),
\end{equation}
with \eqref{eqn:condition_limit_a4_series}.
By using Heine's transformation\cite{Koekoek:book}
\begin{equation}\label{eqn:heine_4}
 {}_2\varphi_1\left(\begin{matrix}a,0\\c\end{matrix};q,z\right)
 =\cfrac{1}{(z;q)_\infty}~
 {}_1\varphi_1\left(\begin{matrix}a^{-1}c\\c\end{matrix};q,az\right),
\end{equation}
$A_2$ is rewritten as
\begin{equation}
 A_2=\cfrac{1}{(p{a_2}^{-1}t;p)_\infty}~
 {}_1\varphi_1\left(\begin{matrix}p{a_2}^{-2}\\-p\end{matrix};p,-pa_2t\right).
\end{equation}
We prepare a suitable integral representation for $A_2$.
\begin{lemma}\label{lemma:1phi1_integral_2}
It holds that
\begin{equation}
 {}_1\varphi_1\left(\begin{matrix}a\\c\end{matrix};q,x\right)
 =\cfrac{(a;q)_\infty(q;q)_\infty}{2\pi i(c;q)_\infty}
  \int_C\cfrac{z^{-\log x/\log q}(cz^{-1};q)_\infty}{(az^{-1};q)_\infty(z;q)_\infty}~dz,
\end{equation}
where the path $C$ runs from $-\infty$ to $+\infty$
so that the poles of $1/(z;q)_\infty$ lie to the right of the path
and the other poles lie to the left of it.
\end{lemma}

\begin{figure}[h]
\begin{center}
\hspace*{9em}
\includegraphics[width=0.9\textwidth]{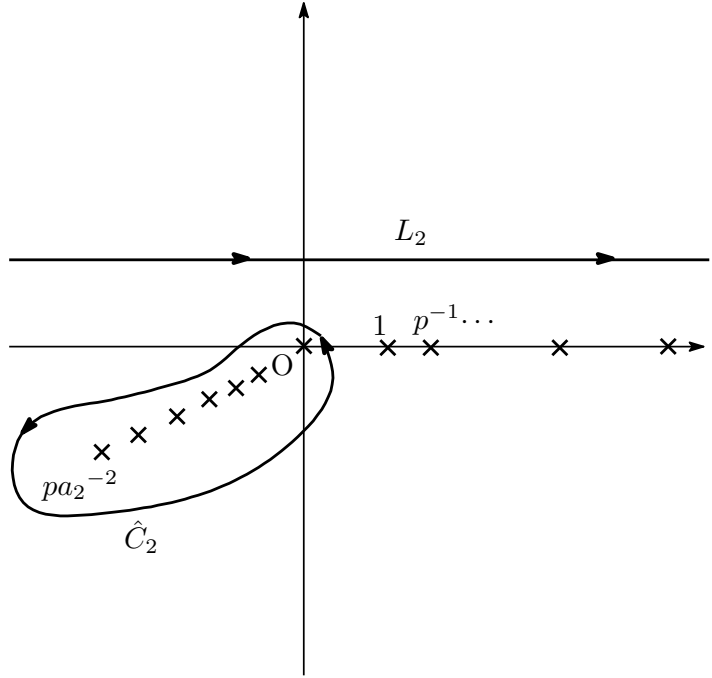}
\end{center}
\caption{The paths of integration $L_2$ and $\hat{C}_2$}
\label{fig:contour_L2_hatC2}
\end{figure}

The Lemma \ref{lemma:1phi1_integral_2} can be proved in a similar manner to Lemma \ref{lemma:1phi1_integral}.
Although Lemma \ref{lemma:1phi1_integral} and Lemma \ref{lemma:1phi1_integral_2} differ only in the integration path, it is appropriate to apply Lemma \ref{lemma:1phi1_integral_2} in order to
choose the integration path in the steepest descent direction.
By using Lemma \ref{lemma:1phi1_integral_2},
it holds that
\begin{equation}
 A_2=\cfrac{(p{a_2}^{-2};p)_\infty(p;p)_\infty}{2\pi i(-p;p)_\infty(p{a_2}^{-1}t;p)_\infty}
  \int_{C_2}\cfrac{z^{-\log (-pa_2t)/\log p}(-pz^{-1};p)_\infty}{(p{a_2}^{-2}z^{-1};p)_\infty(z;p)_\infty}~dz,
\end{equation}
where the path $C_2$ runs from $-\infty$ to $+\infty$
so that the poles of $1/(z;p)_\infty$ lie to the right of the path
and the other poles lie to the left of it.
We divide the path $C_2$ into $L_2$ which runs from $-\infty$ to $+\infty$
so that all poles lie to the right of the path
and $\hat{C}_2$ which is an anticlockwise-oriented contour and encircles 
all poles except for the poles of $1/(z;p)_\infty$ (as shown in Figure \ref{fig:contour_L2_hatC2}).

{\boldmath\bf (1) Path $L_2$.}\quad 
Set
\begin{equation}
 I_3(s)=\int_{L_2}\cfrac{e^{-\log (-pa_2t)\log z/\log p}(-pz^{-1};p)_\infty}{(p{a_2}^{-2}z^{-1};p)_\infty(z;p)_\infty}~dz.
\end{equation}
In a similar manner to $I_1(s)$, one can show that
\begin{equation}
 I_3(s)
 =(-1)^{-3\alpha+1}2^{-\alpha/2}\epsilon^{-\alpha+1} e^{(-\pi^2/12+\pi i\log2)/\epsilon^2-\pi s/\epsilon}
 \int_{\hat{L}_2}  e^{-u^2/2-2^{1/2}isu}u^{-\alpha}[1+O(\epsilon)]du,
\end{equation}
under the assumption
\begin{align}
 &\epsilon>0\quad\text{\rm (when ${\rm Re}(s)<0$)},\\
 &\epsilon<0\quad\text{\rm (when ${\rm Re}(s)>0$)}.
\end{align}
Here the path $\hat{L}_2$ runs from 
$-\infty$ to $+\infty$ when ${\rm Re}(s)>0$ and $+\infty$ to $-\infty$ when ${\rm Re}(s)<0$
so that $u=0$ lies to the right of the path.

{\boldmath\bf (2) Contour $\hat{C}_2$.}\quad
Set 
\begin{equation}\label{eqn:def_I4}
 I_4(s)=\int_{\hat{C}_2}\cfrac{e^{-\log (-pa_2t)\log z/\log p}(-pz^{-1};p)_\infty}{(p{a_2}^{-2}z^{-1};p)_\infty(z;p)_\infty}~dz.
\end{equation}
In the following, we  transform the integral in \eqref{eqn:def_I4} into another suitable  integral, 
then evaluate its asymptotic behavior. 
More precisely,  we first identify $I_4(s)$ with appropriate basic hypergeometric function, 
then apply Heine's transformation and reconstruct the integral representation of it. 
We next evaluate its asymptotic behavior.
\begin{lemma}\label{lemma:2phi1_integral_2}
It holds that
\begin{align}
 &\int_C\cfrac{e^{-\log (-pa_2t)\log z/\log p}(-pz^{-1};p)_\infty}{(p{a_2}^{-2}z^{-1};p)_\infty(z;p)_\infty}~dz\notag\\
 &\quad=-\cfrac{2\pi i(-{a_2}^2;p)_\infty e^{2\log (-pa_2t)\log a_2/\log p}}{{a_2}^3t(p;p)_\infty(p{a_2}^{-2};p)_\infty}~
 {}_2\varphi_1\left(\begin{matrix}p{a_2}^{-2},-p{a_2}^{-2}\\0\end{matrix};p,\cfrac{a_2}{t}\right),
\end{align}
where $C$ is an anticlockwise-oriented contour and encircles 
all poles except for the poles of $1/(z;p)_\infty$.
\end{lemma}

Lemma \ref{lemma:2phi1_integral_2} can be proved in a similar manner to Lemma \ref{lemma:2phi1_integral_1}.
Then $I_4(s)$ is rewritten by using Lemma \ref{lemma:2phi1_integral_2} as
\begin{equation}
 I_4(s)
 =-\cfrac{2\pi i(-{a_2}^2;p)_\infty e^{2\log (-pa_2t)\log a_2/\log p}}{{a_2}^3t(p;p)_\infty(p{a_2}^{-2};p)_\infty}~
 {}_2\varphi_1\left(\begin{matrix}p{a_2}^{-2},-p{a_2}^{-2}\\0\end{matrix};p,\cfrac{a_2}{t}\right).
\end{equation}
Further, by applying Heine's transformation\cite{Koekoek:book}
\begin{equation}\label{eqn:heine_5}
 {}_2\varphi_1\left(\begin{matrix}a,b\\0\end{matrix};q,z\right)
 =\cfrac{(bz;q)_\infty}{(z;q)_\infty}~
 {}_1\varphi_1\left(\begin{matrix}b\\bz\end{matrix};q,az\right),
\end{equation}
and using Lemma \ref{lemma:1phi1_integral_2}, we obtain
\begin{equation}\label{eqn:I4}
 I_4(s)
 =-\cfrac{(-{a_2}^2,-p{a_2}^{-2};p)_\infty e^{2\log (-pa_2t)\log a_2/\log p}}{{a_2}^3t(p{a_2}^{-2},a_2t^{-1};p)_\infty}
  \int_{C_3}\cfrac{e^{-\log (p{a_2}^{-1}t^{-1})\log z/\log p}(-p{a_2}^{-1}t^{-1}z^{-1};p)_\infty}
   {(-p{a_2}^{-2}z^{-1};p)_\infty(z;p)_\infty}~dz,
\end{equation}
where the path $C_3$ runs from $-\infty$ to $+\infty$
so that the poles of $1/(z;p)_\infty$ lie to the right of the path
and the other poles lie to the left of it.
We finally evaluate the integral of the right hand side of (5.77). 
To this end, we divide the path $C_3$ into $L_3$ which runs from $-\infty$ to $+\infty$
so that all poles lie to the right of the path
and $\hat{C}_3$ which is an anticlockwise-oriented contour and encircles
all poles except for the poles of $1/(z;p)_\infty$.

Set
\begin{align}
 &I_4^{(1)}(s)=\int_{L_3}\cfrac{e^{-\log (p{a_2}^{-1}t^{-1})\log z/\log p}(-p{a_2}^{-1}t^{-1}z^{-1};p)_\infty}
   {(-p{a_2}^{-2}z^{-1};p)_\infty(z;p)_\infty}~dz,\\
 &I_4^{(2)}(s)=\int_{\hat{C}_3}\cfrac{e^{-\log (p{a_2}^{-1}t^{-1})\log z/\log p}(-p{a_2}^{-1}t^{-1}z^{-1};p)_\infty}
   {(-p{a_2}^{-2}z^{-1};p)_\infty(z;p)_\infty}~dz.
\end{align}
In a similar manner to $I_1(s)$,
the asymptotic behavior of $I_4^{(1)}(s)$ as $\epsilon \to 0$ can be evaluated as
\begin{equation}
 I_4^{(1)}(s)
 =(-1)^{1/2+\alpha}2^{-3\alpha/2}\epsilon^{1-\alpha} e^{(\pi^2+2(\log2)^2)/4\epsilon^2+is\log2/\epsilon+s^2/2}
 \int_{\hat{L}_2}e^{-u^2/2-2^{1/2}isu} u^{-\alpha}[1+O(\epsilon)]~du.
\end{equation}
$I_4^{(2)}(s)$ can be expressed in terms of the basic hypergeometric function.  
In fact, we have the following lemma which is proved in a similar manner to Lemma \ref{lemma:2phi1_integral_1}:
\begin{lemma}\label{lemma:2phi1_integral_3}
It holds that
\begin{align}
 &\int_C\cfrac{e^{-\log (p{a_2}^{-1}t^{-1})\log z/\log p}(-p{a_2}^{-1}t^{-1}z^{-1};p)_\infty}
   {(-p{a_2}^{-2}z^{-1};p)_\infty(z;p)_\infty}~dz\notag\\
 &\quad=-\cfrac{2\pi it(a_2t^{-1};p)_\infty e^{-\log (p{a_2}^{-1}t^{-1})\log (-{a_2}^{-2})/\log p}}{a_2(-p{a_2}^{-2},p;p)_\infty}~
 {}_2\varphi_1\left(\begin{matrix}p{a_2}^{-1}t,-p{a_2}^{-2}\\0\end{matrix};p,{a_2}^2\right),
\end{align}
where $C$ is an anticlockwise-oriented contour and encircles
all poles except for the poles of $1/(z;p)_\infty$.
\end{lemma}

Then by using Lemma \ref{lemma:2phi1_integral_3}, $I_4^{(2)}(s)$ yields
\begin{equation}
 I_4^{(2)}(s)=-\cfrac{2\pi it(a_2t^{-1};p)_\infty e^{-\log (p{a_2}^{-1}t^{-1})\log (-{a_2}^{-2})/\log p}}{a_2(-p{a_2}^{-2},p;p)_\infty}~
 {}_2\varphi_1\left(\begin{matrix}p{a_2}^{-1}t,-p{a_2}^{-2}\\0\end{matrix};p,{a_2}^2\right),
\end{equation}
which is rewritten  by applying Heine's transformation \eqref{eqn:heine_3} as
\begin{equation}
 I_4^{(2)}(s)=-\cfrac{2\pi it(a_2t^{-1},-p,p{a_2}^{-1}t;p)_\infty e^{-\log (p{a_2}^{-1}t^{-1})\log (-{a_2}^{-2})/\log p}}
  {a_2(-p{a_2}^{-2},p,{a_2}^2;p)_\infty}~
 {}_2\varphi_1\left(\begin{matrix}{a_2}^2,0\\-p\end{matrix};p,p{a_2}^{-1}t\right).
\end{equation}
Therefore we finally obtain
\begin{align}
 A_2
 =&\cfrac{(p{a_2}^{-2};p)_\infty(p;p)_\infty}{2\pi i(-p;p)_\infty(p{a_2}^{-1}t;p)_\infty}
  \left(I_3(s)
  -\cfrac{(-{a_2}^2,-p{a_2}^{-2};p)_\infty e^{2\log (-pa_2t)\log a_2/\log p}}{{a_2}^3t(p{a_2}^{-2},a_2t^{-1};p)_\infty}
  \left(I_4^{(1)}(s) + I_4^{(2)}(s)\right)
  \right)\\
 =&(-1)^{-1-4\alpha}2^{-1/2+\alpha/2}\pi^{-1/2}\epsilon^{-\alpha} e^{s^2+2\pi i\log2/\epsilon^2}
   \int_{L}  e^{-u^2/2-2^{1/2}isu}u^{-\alpha}[1+O(\epsilon)]du\notag\\
  &+\cfrac{(-{a_2}^2;p)_\infty e^{(2\log (-pa_2)\log a_2-\log (p{a_2}^{-1})\log (-{a_2}^{-2}))/\log p}}
    {{a_2}^4({a_2}^2;p)_\infty}~(-1)^{\log t/\log p}
  {}_2\varphi_1\left(\begin{matrix}{a_2}^2,0\\-p\end{matrix};p,p{a_2}^{-1}t\right),
\end{align}
where we have used Proposition \ref{prop:qshifted_asymptotic}.
This proves Lemma \ref{lemma:continuous_weber_2}.
\section{Concluding remarks}
In this paper, we have constructed the hypergeometric solutions to the symmetric $q$-Painlev\'e
equations of the types $(A_2+A_1)^{(1)}$, $A_4^{(1)}$, $D_5^{(1)}$, $E_6^{(1)}$, $E_7^{(1)}$ 
and $E_8^{(1)}$, and discussed their continuous limits.
In particular, we have shown that the hypergeometric function appearing in the solution
to the symmetric $q$-P$(A_4^{(1)})$ actually reduces to the Weber function by applying
the saddle point method to its integral representation.

Before closing, we give a remark on a $q$-Painlev\'e equation of the type $A_4^{(1)}$.
The following $q$-difference equation
\begin{equation}\label{eqn:a4_sakai}
 \overline{g}g=\cfrac{(f+t^{-1})(f+\alpha t^{-1})}{1+\gamma f},\quad
 f\underline{f}
 =\cfrac{(g+q^{1/2}\alpha\beta t^{-1})
  (g+q^{1/2}{\beta}^{-1}t^{-1})}{1+{\gamma}^{-1}g},
\end{equation}
is usually referred to as a $q$-Painlev\'e equation of type $A_4^{(1)}$.
Equation \eqref{eqn:a4_sakai} also describes a translation on the root lattice 
of type $A_4^{(1)}$, but its direction is different from that of \eqref{eqn:a4}.
It may be interesting to note that the symmetric $q$-Painlev\'e equation
obtained from \eqref{eqn:a4_sakai} by the projective reduction
\begin{equation}\label{eqn:sym_a4_sakai}
 \widetilde{X}\wutilde{X}=\cfrac{(X+t^{-1})(X+\alpha t^{-1})}{1+X},
\end{equation}
has no hypergeometric solution (see \cite{HKNN:qP3D7}) which is a sharp contrast to \eqref{eqn:sym_a4}.
\par\bigskip
\noindent{\bf Acknowledgement.} 
The authors would like to express their sincere thanks to
Mr. P. Howes, Prof. H. Ochiai and Prof. T. Tsuda for fruitful discussions and valuable suggestions. 
This work has been partially
supported by JSPS Grant-in-Aid for Scientific Research No. 23340037 and 22$\cdot$4366.


\end{document}